\documentclass[usenatbib,a4paper,times,fleqn]{mnras}
\usepackage[T1]{fontenc}
\usepackage{graphicx,natbib,times,amssymb,amsmath,pstricks,mathtools,enumitem,array}
\usepackage{aas_macros}
\usepackage{bm,url}
\usepackage{textcomp}
\usepackage{breakurl}
\usepackage[latin1]{inputenc}

\usepackage{nicefrac,xfrac}
\usepackage{mathtools}
\usepackage{cancel}
\usepackage{listings}
\usepackage{pifont}
\usepackage{parskip}
\usepackage{siunitx}
\usepackage{booktabs}
\usepackage[font=Large]{subfig}
\usepackage[normalem]{ulem}

\usepackage{cuted}

\newcommand\redsout{\bgroup\markoverwith{\textcolor{red}{\rule[0.5ex]{2pt}{1pt}}}\ULon}
\newcommand {\apgt} {\ {\raise-.5ex\hbox{$\buildrel>\over\sim$}}\ }
\newcommand {\aplt} {\ {\raise-.5ex\hbox{$\buildrel<\over\sim$}}\ }

\title[Fragmentation with Discontinuous Galerkin schemes]{Fragmentation with Discontinuous Galerkin schemes: Non-linear fragmentation}

\author[Lombart \& Hutchison \& Lee]{Maxime Lombart$^{1}$\thanks{maxime.lombart@gapps.ntnu.edu.tw}, Mark Hutchison$^{2}$, Yueh-Ning Lee$^{1}$  \\
$^{1}$Department of Earth Sciences, National Taiwan Normal University, 88, Sec.4, Ting-Chou Road, Taipei 11677, Taiwan\\
$^{2}$Universit{\"a}ts-Sternwarte, Ludwig-Maximilians-Universit{\"a}t  M{\"u}nchen, Scheinerstr. 1, 81679 M{\"u}nchen, Germany
}
\date{}
\begin{document}
%\label{firstpage}
\bibliographystyle{mnras}
\maketitle

\begin{abstract}
Small grains play an essential role in astrophysical processes such as chemistry, radiative transfer, gas/dust dynamics. The population of small grains is mainly maintained by the fragmentation process due to colliding grains. An accurate treatment of dust fragmentation is required in numerical modelling. However, current algorithms for solving fragmentation equation suffer from an over-diffusion in the conditions of 3D simulations. To tackle this challenge, we developed a Discontinuous Galerkin scheme to solve efficiently the non-linear fragmentation equation with a limited number of dust bins. \\
\end{abstract}


%----------------------------------------------------------------------------------------------------------------
\section{Introduction}
\label{sec:introduction}
Fragmentation resulting from collisions between grains is ubiquitous in astrophysics at all scales: asteroid belts \citep{Williams1994,Bottke2005}, debris discs \citep{Kenyon2004,Kobayashi2010a}, protoplanetary discs \citep{Safronov1972,Birnstiel2016,Blum2018}, planetary rings \citep{Brilliantov2015} and molecular clouds \citep{Draine2004, Jones2004}. Fragmentation counter-balances the coagulation process by forming small grains and maintaining a polydisperse dust size distribution. The role of dust grains is fundamental in many astrophysical processes: thermal emission \citep[e.g.][]{Draine2004, Andrews2020}, formation of H$_2$ \citep[e.g.][]{Gould1963,Jones2021}, dynamics \citep[e.g.][]{Testi2014,Lesur2022}. Specifically large grains tend to decouple dynamically from the gas. For example in star formation, large grains preferentially accumulate in regions of high gas density \citep{Lebreuilly2021}. In protoplanetary discs, large grains drift radially inwards towards the star after losing momentum to the gas through drag \citep[][and references therein]{Whipple1972,Weidenschilling1977,Lesur2022}. Meanwhile, the acceleration of the gas by the dust, or "backreaction", is central to the operation of the streaming instability \citep[][and references therein]{Youdin2005,Testi2014,Jaupart2020,Lesur2022} and, consequently, planetesimal formation \citep{Gonzalez2017}.  All these physical processes emphasise the importance of accounting for the evolution of the dust size distribution at all scales in astrophysics. \\

Key to understanding the impact of dust on the interstellar medium (ISM) and protoplanetary discs is the knowledge of how the complete dust size distribution evolves in time. This study focuses on the non-linear fragmentation \citep{Cheng1990,Kostoglou2000,Ernst2007,Pagonabarraga2009} that occurs when two grains collide, which plays an important role in maintaining the small grain population in astrophysical environments. This fragmentation model is distinct from the linear fragmentation model where the breakup of grains is driven spontaneously by external forces where dust collisions are rare \citep{Cheng1990,Kostoglou2000}. Details of this model are given in Appendix \ref{ap:lin_frag}. The non-linear fragmentation models the breakup of grains of any size.
\begin{figure}
\includegraphics[width=0.9\columnwidth]{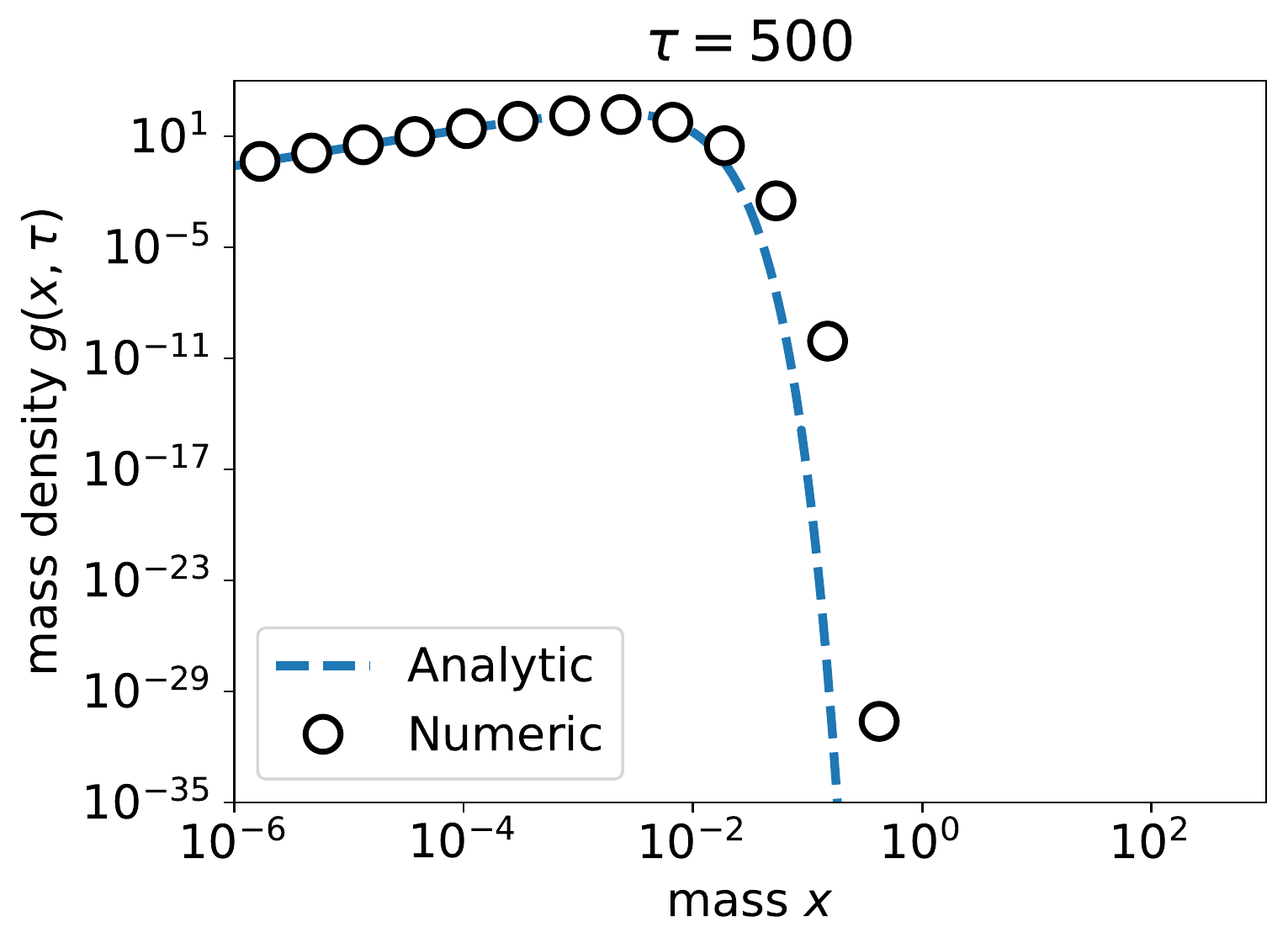}
\caption{Numerical diffusion problem for fragmentation: numerical schemes of order 0 over-estimate the mass density of large grains at low resolution, here for the case of a non-linear fragmentation equation with multiplicative collision kernel and $N=20$ logarithmically-spaced dust bins.}
\label{fig:overdiff} 
\end{figure}
The non-linear fragmentation process is formalised within the framework of Smoluchowski-like equation by the mean-field non-linear fragmentation equation \citep{Redner1990,Kostoglou2000,Ernst2007,Pagonabarraga2009,DaCosta2015,Banasiak2019}. The non-linear fragmentation equation does not have generic analytic solutions but its mathematical description is linked to the linear fragmentation equation \citep[e.g.][]{Redner1990,Kostoglou2000,Ernst2007,Pagonabarraga2009,Paul2018,Barik2018,Das2020}. For astrophysics problems, the non-linear fragmentation equation has to be solved numerically.\\

Dust evolution (i.e. coagulation and fragmentation) is important at many different scales in astrophysics, from the ISM to planet formation \citep[e.g.][]{Safronov1972,Tanaka1996,Bottke2005,Brauer2008,Hirashita2009,Ormel2009,Kobayashi2010a,Brilliantov2015,Blum2018,Jones2021}. Tracking the size evolution of dust in 3D simulations of these environments would be a big step forward, but is currently hindered by the lack of numerical schemes that are both economical and accurate -- particularly in the case of non-linear fragmentation where large grains fragments into small grains. For example, Fig. \ref{fig:overdiff} illustrates the over-diffusion problem for fragmentation commonly seen numerical schemes of order 0 with insufficient mass resolution (shown here with 20 bins). The mass density of large grains becomes increasingly inaccurate with time, which later impacts the evolution of small grains. Brute force has been the standard approach to combating over diffusion in the past -- typically employing hundreds of mass bins to maintain accuracy. In comparison, 3D hydrodynamical simulations can feasibly only accommodate a few tens of bins, making the two methods incompatible. Inspired by the recent work of \citet{Liu2019} and \citet{Lombart2021}, we took a different approach to modelling the non-linear fragmentation equation that significantly closes this computation gap. Using a high-order solver based on the Discontinuous Galerkin method we found we could reduce the number of mass bins without sacrificing accuracy. Our goal in this paper is to apply these same methods to the non-linear fragmentation equation.\\

The structure of the paper is as follows: Properties of the non-linear fragmentation equation discussed in the astrophysical context are presented in Sect.~\ref{sec:frag}. The Discontinuous Galerkin numerical scheme is presented in Sect.~\ref{sec:dg}. The performance of the solver regarding the over-diffusion problem are studied in Sect.~\ref{sec:num}. Applicability of the algorithm in astrophysical and other contexts are discussed in Sect.~\ref{sec:discussions}.

%-----------------------------------------------------------------------------------------------------------------
\section{Non-linear fragmentation equation}
\label{sec:frag}

The non-linear fragmentation equation describes the mass distribution of grains where mass transfers from large to small grains. The fragmentation process is modelled by a non-linear partial integro-differential hyperbolic equation that depends on two quantities: i) the fragmentation kernel, which describes the collision rate between two grains, and ii) the distribution of outgoing fragments. Specifically, this non-linear fragmentation model considers that only one of the two colliding grains breaks-up (see Sect.~\ref{sec:dist_frag}), which is the case for a collision between a large and a small grains. The non-linear fragmentation equation was initially formalised in fields of chemistry and atmospheric science \citep[][and references therein]{Kostoglou2000}. Explicit solutions exist only for simple forms of the fragmentation kernel, such as the constant and multiplicative kernels, and the distribution of fragments \citep{Ziff1985,McGrady1987,Kostoglou2000,Ernst2007}. Many of the analytic solutions are similar to the linear case \citep{Kostoglou2000} and share properties, such as self-similarity and shattering \citep{McGrady1987,Cheng1988,Cheng1990,Kostoglou2000,Ernst2007}. In astrophysics, collisions leading to fragmentation occur mainly through ballistic impacts \citep{Safronov1972,Tanaka1996,Dullemond2005,Kobayashi2010a}, for which no analytical solutions exist and require numerical solutions.

\subsection{Conservative form}
\label{sec:cons_form}
\begin{figure}
\includegraphics[width=\columnwidth,trim=80 150 50 50, clip]{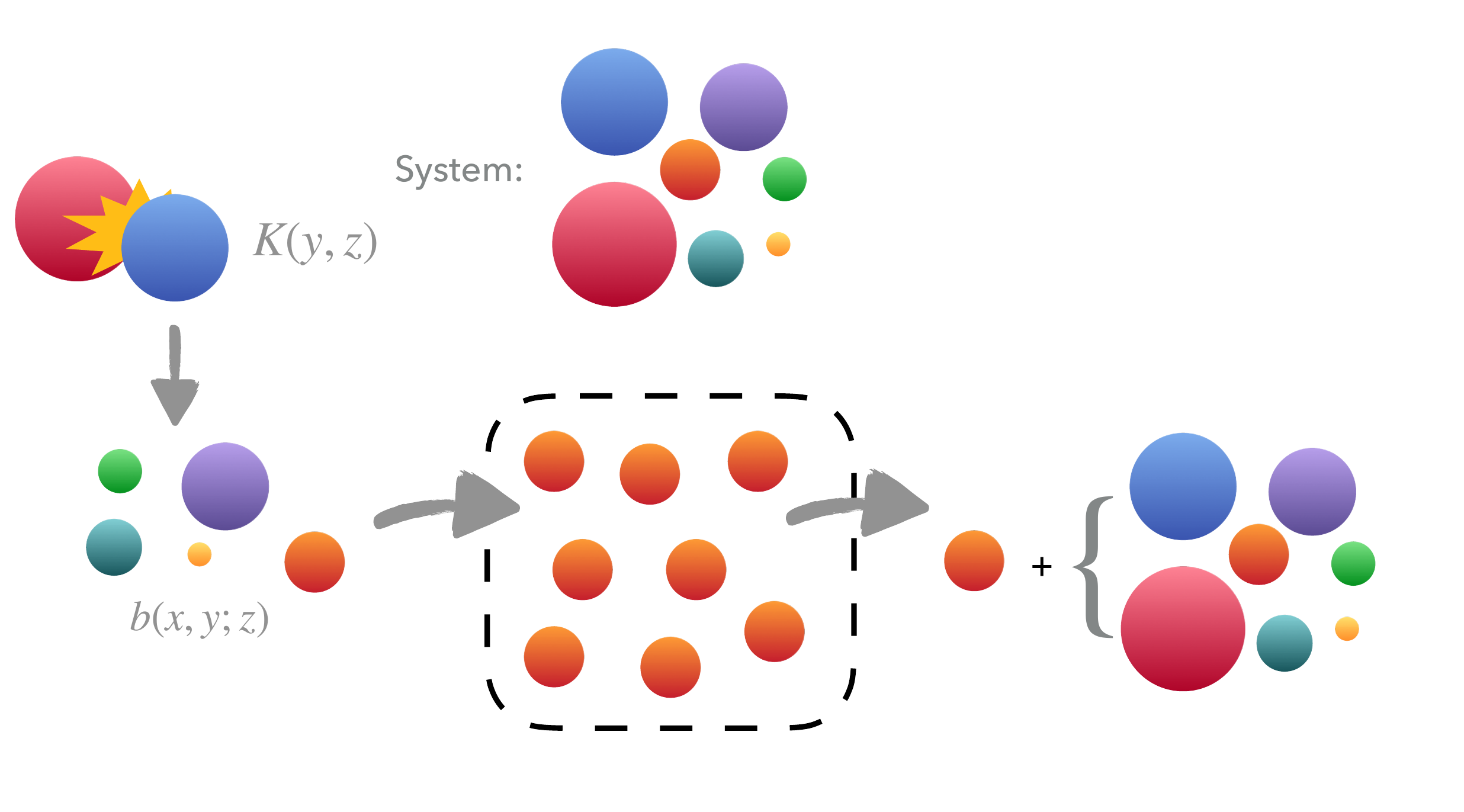}
\caption{Scheme of the collisional fragmentation equation Eq.~\ref{eq:frag_cont}. The number density of orange grains, mass $m$ increases du to fragmentation of the blue grains with mass greater than $m$, and decreases due to collision between an orange grain and any grain in the system.}
\label{fig:scheme_frag}
\end{figure}

The non-linear fragmentation equation was formalised in a mean-field approach by \citet{Kostoglou2000}.
Fragments are assumed to be spherical. Spatial correlations are neglected. For physical systems involving the fragmentation of aggregates made of a large number of monomers, it is convenient to assume a continuous mass distribution. The population density of grains within an elementary mass range $\mathrm{d}m$ is characterised by its number density $n(m)$. The continuous collisional fragmentation equation is given by
\begin{equation}
\begin{aligned}
& \frac{\partial n (m,t)}{\partial t } = \\
& \qquad \int\limits_{0}^{\infty} \int\limits_{m}^{\infty} K(m',m'') b(m,m';m'') n(m',t) n(m'',t) \mathrm{d}m' \mathrm{d}m'' \\
& \qquad \qquad - n(m,t) \int\limits_{0}^{\infty} K(m,m') n(m',t) \mathrm{d}m', 
\end{aligned}
\label{eq:frag_cont}
\end{equation}
where $t$ denotes time, $m'$ and $m''$ the masses of two colliding grains, $m$ the mass of fragments and $n(m,t)$ is the number density function by mass unit of particles of mass $m$. The first term on the right-hand side of Eq.~\ref{eq:frag_cont} represents the increase of particle of mass $m$ produced by the fragmentation of a particle of mass $m'$ due to collision with a particle of mass $m''$ (breaking of a large particle to form an orange particle in Fig.~\ref{fig:scheme_frag}). The averaged probabilities of collision leading to fragmentation are encoded inside the fragmentation kernel $K(m',m'')$, which is a symmetric function of $m'$ and $m''$ for binary collisions.  The second term of Eq.~\ref{eq:frag_cont} accounts for the loss of particles of mass $m$ due to their fragmentation into smaller particles (right side on Fig.~\ref{fig:scheme_frag}), the function $b(m,m';m'')$ being the distribution of outgoing fragments of mass $m$.

If $n_0(m) = n(m,0)$, is the initial number density distribution per unit mass, then the total mass density, the total number density of grains and the mean mass of the initial distribution can be written as
\begin{equation}
M = \int\limits_0^{\infty} mn_0(m) \mathrm{d}m, \, N_0 = \int\limits_0^{\infty} n_0(m) \mathrm{d}m,\, m_0=\frac{M}{N_0}.
\end{equation}
The non-linear fragmentation equation writes in dimensionless form by using \citep{Kostoglou2000}
\begin{equation}
  \left\{ 
  \begin{aligned}
    & x \equiv m/m_0,\,y \equiv m'/m_0,\, \mathcal{K}(x,y) = K(m,m')/K(m_0,m_0), \\
    & \tau = (K(m_0,m_0) N_0) t,\, f(x,\tau) = m_0 \, n(m,t)/N_0,\\
    & \tilde{b}(x,y;z) = m_0 b(m,m';m'').
  \end{aligned}
  \right.
\end{equation}
Here, $K(m_0,m_0)$ is a normalising constant with dimensions $\rm [length]^3 \; [time]^{-1}$. We use the variables $x$, $\tau$, and $f$ for the dimensionless mass, time, and number density, respectively, to be consistent with the existing literature \citep[e.g.][]{Kostoglou2007,Banasiak2019} such that Eq.~\ref{eq:linfrag_cont} writes
\begin{equation}
\begin{aligned}
\frac{\partial f (x,\tau)}{\partial \tau} = & \int\limits_{0}^{\infty} \int\limits_{x}^{\infty} \mathcal{K}(y,z) \tilde{b}(x,y;z) f(y,\tau) f(z,\tau) \mathrm{d}y \mathrm{d}z \\
& - f(x,\tau) \int\limits_{0}^{\infty} \mathcal{K}(x,y) f(y,\tau) \mathrm{d}y.
\end{aligned}
\label{eq:frag_cont_DL}
\end{equation}
Recently, \citet{Paul2018} have shown that Eq.~\ref{eq:frag_cont_DL} can be equivalently written in the conservative form
\begin{equation}
\left\{
   \begin{aligned}
   &\frac{\partial g \left( x,\tau \right) }{\partial \tau} - \frac{\partial F_{\mathrm{frag}} \left[ g \right] \left( x,\tau \right)}{\partial x} = 0,\\
   &F_{\mathrm{frag}} \left[ g \right] \left( x,\tau \right) \equiv \\
   & \int\limits_0^{\infty} \int\limits_x^{\infty} \int\limits_0^x \frac{w}{uv} \tilde{b}(w,u;v) \mathcal{K}(u,v) g(v,\tau) g(u,\tau)  \mathrm{d}w \mathrm{d}u \mathrm{d}v,
   \end{aligned}
\right.
\label{eq:frag_cons_DL}
\end{equation}
where $g(x,\tau) \equiv xf(x,\tau)$ is the mass density of grains per unit mass, and $F_{\mathrm{frag}} [g] (x,\tau)$ is the flux of mass density across the mass $x$ triggered by fragmentation.

\subsection{Kernels}
\label{sec:kernels}
The collision kernel is the same as the one used in coagulation theory \citep{Hidy1972,Kostoglou2000,Ernst2007}. The restriction on the form of $\mathcal{K}$ is that it has to be a symmetric function, i.e.n $\mathcal{K}(x,y) =  \mathcal{K}(y,x)$. Exact solutions of the non-linear fragmentation equation are summarised in Table~\ref{table:kernels} for constant and multiplicative kernels. A constant kernel means that collision frequency is independent of the size of colliding grains. This is the case for an initial monodispersed system of grains for a short time period \citep[][and references therein]{Hidy1972}. No physical interpretation can be given for the multiplicative kernel since, at least in the case of coagulation, it leads to a singularity in the Smoluchowski coagulation equation \citep{Leyvraz1981}. In general, these simple kernels are only used to benchmark algorithms with their associated exact solution. The preferred fragmentation kernel used in astrophysics is the ballistic kernel described in Sect.~\ref{sec:discussions}.

\begin{table*}
\begin{center}
\begin{tabular}{ccc}
  \hline
  Kernel & $\mathcal{K}(x,y)$ & exact solution\\
  \hline
  Constant & $1$  & $f(x,\tau') = f(x,0)\exp(-\tau') + \sqrt{2 \tau'} \exp(-\tau') \int\limits_x^{\infty} \frac{I_1\left( 2 \sqrt{2\tau' \log(y/x)} \right) f(y,0)}{y \sqrt{\log(y/x)}} \mathrm{d}y$ \\
  Multiplicative& $xy$ & $f(x,\tau) = (1+\tau)^2 \exp(-x (1+\tau))$ \\
\end{tabular}
\caption{The two fragmentation kernels $\mathcal{K}$ considered in this study with the corresponding exact solution.}
\label{table:kernels}
\end{center}
\end{table*}

\subsection{Distribution of fragments}
\label{sec:dist_frag}
The function $\tilde{b}(x,y;z)$ is the probability density function for the formation of particle of mass $x$ resulting from the break-up of particle of mass $y$ due to its collision with a particle of mass $z$. The function $\tilde{b}(x,y;z)$  is an important physical quantity in the fragmentation process that determines the distribution of outgoing fragments of mass $x$  \citep{Cheng1990,Kostoglou2000,Banasiak2019}. The $z$ dependence in the function $\tilde{b}(x,y;z)$ reflects that a grain of mass $y$ will give a different fragment distribution depending on the mass of the impactor. The function $\tilde{b}(x,y;z)$ satisfies the following requirements
\begin{equation}
\tilde{b}(x,y;z) = 
\begin{cases}
\tilde{b}(x,y;z) &x,y,z >0\\
0 & \forall x>y
\end{cases}
\end{equation}
and local conservation of mass
\begin{equation}
\forall y>0,\; \int\limits_0^y x \tilde{b}(x,y;z) \mathrm{d}x = y,
\label{eq:cons_mass}
\end{equation}
such that the total mass of the fragments is equal to the mass of the parent particle $y$. According to Eq.~\ref{eq:cons_mass}, it is important to note that, after the collision, only particle $y$ breaks up while $z$ remains intact. According to the definition of $\tilde{b}$, the number of fragments writes 
\begin{equation}
\tilde{N}(y;z) = \int\limits_0^y \tilde{b}(x,y;z) \mathrm{d}x.
\end{equation}
Moreover, $\tilde{b}$ should satisfy the following requirement \citep{Kostoglou2000,Banasiak2019}
\begin{equation}
\int\limits_0^k x \tilde{b}(x,y;z) \mathrm{d}x \geq \int\limits_{y-k}^y (y-x) \tilde{b}(x,y;z) \mathrm{d}x,\; 0 \leq k \leq y/2.
\label{eq:condition_b}
\end{equation}
The inequality in Eq.~\ref{eq:condition_b} ensures that when a fragment $x \ge y/2$ is formed, the mass from the remaining fragments between $x$ and $y$ (right-hand side) contributes to, but does not exceed, the total mass of grains between 0 and $k = y-x$ (left-hand side) that could additionally include some pre-existing (i.e. unbroken) grains. This ensures that no rearrangement of the mass is allowed.

\subsection{Analytic solutions}
\label{sec:analytic}
The number density distribution that results from Eq.~\ref{eq:frag_cont_DL} is determined by the details of the collision kernel $\mathcal{K}(x,y)$ and the distribution of fragments $\tilde{b}(x,y;z)$. Exact solutions are known only for simple forms of $\mathcal{K}$ (e.g. the constant and multiplicative kernels) and for homogeneous forms of $\tilde{b}$. Homogeneity implies that the distribution of fragments $\tilde{b}$ can be written as a power law
\begin{equation}
\tilde{b}(x,y;z) =  p(x/y)/y. 
\end{equation}
Due to the restrictions from Eqs.~\ref{eq:cons_mass} and \ref{eq:condition_b}, $p$ takes the form \citep{McGrady1987,Cheng1990,Ernst2007}
\begin{equation}
p(x) = (\beta +2) x^{\beta},
\label{eq:p}
\end{equation}
obeying,
\begin{equation}
\int\limits_0^y x \tilde{b}(x,y;z) \mathrm{d}x = y,\; \tilde{N}(y) = \frac{\beta+2}{\beta+1}.
\label{eq:mass_cons}
\end{equation}
By definition, fragmentation results in at least two fragments. Therefore, the number of fragments satisfies $\tilde{N} \geq 2$ which implies $-1 < \beta \leq 0$. Exact solutions have been derived for the constant and multiplicative collision kernels in the case of binary breakage ($\beta = 0$) where $\tilde{b}(x,y;z) = 2/y$ \citep{Ziff1985,Kostoglou2000}. Existence and uniqueness of the following exact solutions for the non-linear fragmentation equation have been proved recently \citep{Paul2018,Barik2018}. We review these solutions (see Table~\ref{table:kernels}) and they will be used to benchmark the algorithm in Sect.~\ref{sec:num}.

\subsubsection{Constant kernel}
\label{subsec:kconst}
By substituting $\mathcal{K}(x,y)=1$ into Eq.~\ref{eq:frag_cont_DL}, we obtain
\begin{equation}
\frac{\partial f (x,\tau)}{\partial \tau} = -N f(x,\tau) + N \int_x^{\infty} \frac{p(x/y)}{y} f(y,\tau) \mathrm{d}y,
\label{eq:kconst_init}
\end{equation}
where $N = \int_0^{\infty} f(x,\tau)$ is the dimensionless total number density of particles. By integrating Eq.~\ref{eq:kconst_init} from $x=0$ to $x=\infty$, the equation on $N$ writes
\begin{equation}
\frac{\mathrm{d} N}{\mathrm{d}\tau}  = (b_0-1)N^2 \; \Rightarrow N(\tau) = \frac{1}{1-(b_0-1)\tau},
\label{eq:kconst_N}	
\end{equation}
where $b_0=\int_0^1 p(w) \mathrm{d}w$, with $w=x/y$. Details are given in Appendix~\ref{ap:kernel_const}. An important remark is that there is a shattering phenomenon at $\tau_c = 1/(b_0-1)$ where the total number of particles diverges. The system no longer conserves mass as particles approach infinitely small masses, analogous to the gelation phenomenon for coagulation in the opposite limit \citep{Ziff1985,Kostoglou2000,Ernst2007}. Then, by using the following change of variables for time
\begin{equation}
\tau' = \int_0^{\tau} N(t) \mathrm{d}t = \frac{1}{1-b_0} \log(1-(b_0-1)\tau),
\label{eq:time_var}
\end{equation}
we obtain 
\begin{equation}
\frac{\partial f (x,\tau')}{\partial \tau' } = - f(x,\tau') +  \int\limits_x^{\infty} \frac{1}{y} p(x/y) f(y,\tau') \mathrm{d}y.
\label{eq:kconst}
\end{equation}
This is identical to the linear fragmentation equation with constant fragmentation rate \citep{Ziff1985,Kostoglou2000}. The solution to Eq.~\ref{eq:kconst} with a general initial condition and $p(x/y)=2$ writes \citep{Ziff1985}
\begin{equation}
\begin{aligned}
& f(x,\tau') =  f(x,0)\exp(-\tau') \\
&  + \sqrt{2 \tau'} \exp(-\tau') \int\limits_x^{\infty} \frac{I_1\left( 2 \sqrt{2\tau' \log(y/x)} \right) f(y,0)}{y \sqrt{\log(y/x)}} \mathrm{d}y,
\end{aligned}
\label{eq:sol_kconst}
\end{equation}
where $I_1$ is the modified Bessel function of first kind. For the benchmark of the numerical scheme in Sect.~\ref{sec:num}, the analytical solution is evaluated with the function \texttt{integrate} from the \textsc{Python} library \texttt{scipy}. 

\subsubsection{Multiplicative kernel}
\label{subsec:kmul}
The solution for the multiplicative kernel $\mathcal{K}(x,y)=xy$ and the initial condition $f_0(x,0)=\exp(-x)$ has been derived by \citep{Ziff1985,Kostoglou2000}. Eq.~\ref{eq:frag_cont_DL} can be written as
\begin{equation}
\frac{\partial f (x,\tau)}{\partial \tau } = - x f(x,\tau) + 2 \int\limits_x^{\infty} f(y,\tau) \mathrm{d}y.
\label{eq:kmul}
\end{equation}
This is identical to the linear fragmentation equation with a linear fragmentation rate \citep{Ziff1985,Kostoglou2000}. The solution of Eq.~\ref{eq:kmul} writes
\begin{equation}
f(x,\tau) = (1+\tau)^2 \exp(-x (1+\tau)).
\label{eq:sol_kmul}
\end{equation}

%-----------------------------------------------------------------------------------------------------------------
\section{Discontinuous Galerkin algorithm}
\label{sec:dg}
\citet{Lombart2021} showed that the discontinuous Galerkin (DG) method could be used to efficiently solve the Smoluchowski coagulation equation with a reduced number of mass bins. We now want to apply the same method to non-linear fragmentation described in Eq.~\ref{eq:frag_cont_DL}. We refer the reader to \citet{Lombart2021} for a complete description of the general algorithm and here only detail the steps that are altered to model the non-linear fragmentation equation, i.e. the evaluation of the flux, the integral of the flux and the Courant-Friedrichs-Lewy condition (CFL). The algorithm flowchart is given in the github repository in Sect.\ref{data_github}.

\subsection{Summary}
Since the non-linear fragmentation Eq.~\ref{eq:frag_cont_DL} is defined over an unbounded mass interval $x \in \mathbb{R}_+$, the first step is to reduce the mass interval to a physically relevant mass range $[x_{\mathrm{min}}>0, x_{\mathrm{max}}<+\infty]$, and divided into $N$ bins. Each bin is defined by $I_j = [x_{j-1/2},x_{j+1/2})$ for $j \in [\![1,N]\!]$. The size and the center position of each bin are defined respectively by $h_j = x_{j+1/2}-x_{j-1/2}$ and $x_j=(x_{j+1/2}+x_{j-1/2})/2$. Eq.~\ref{eq:frag_cons_DL} is then multiplied by a Legendre polynomial basis function $\phi$ and integrated over each bin. After using an integral by part transformation, the DG method consist in solving the following equation
\begin{equation}
\begin{aligned}
\int_{I_j} \frac{\partial g_j}{\partial x} \phi \mathrm{d}x &+ \underbrace{\int_{I_j} F_{\mathrm{frag}}[g](x,\tau) \frac{\partial \phi}{\partial x} \mathrm{d}x}_{\text{integral of the flux see Sect.~\ref{sec:intflux}}} \\
&   - F_{\mathrm{frag}}[g](x_{j+1/2},\tau) \phi(x_{j+1/2}) \\
&  + F_{\mathrm{frag}}(x_{j-1/2},\tau) \phi(x_{j-1/2}) = 0,
\end{aligned}
\label{eq:DG}
\end{equation}
where $g_j$ is the Legendre polynomials approximation of $g$ in bin $I_j$
\begin{equation}
\forall x \in I_j,\; g(x) \approx g_j(x,\tau) = \sum_{i=0}^k g_j^i(\tau) \phi_i(\xi_j(x)).
\end{equation}
The function $\xi$ map the bin interval into interval $[-1,1]$ and $k$ is the order of the Legendre polynomials.

\subsection{Numerical flux evaluation}
\label{sec:fluxes}
The truncating fragmentation flux into the physically relevant mass range $[x_{\mathrm{min}}>0, x_{\mathrm{max}}<+\infty]$ writes
\begin{equation}
\begin{aligned}
&F_{\mathrm{frag}} \left[ g \right] \left( x,\tau \right) = \\
& \int\limits_{x_{\mathrm{min}}}^{x_{\mathrm{max}}} \int\limits_x^{x_{\mathrm{max}}} \int\limits_{x_{\mathrm{min}}}^x \frac{w}{uv} \tilde{b}(w,u;v) \mathcal{K}(u,v) g(v,\tau) g(u,\tau)  \mathrm{d}w \mathrm{d}u \mathrm{d}v.
\end{aligned}
\label{eq:frag_flux}
\end{equation}
Mass conservation is preserved by not allowing particles of mass larger than $x_{\mathrm{max}}$ or smaller than $x_{\mathrm{min}}$ to form.

\subsubsection{Description of the flux}
The fragmentation flux $F_{\mathrm{frag}}$ is non local, similar to the coagulation flux \citep{Liu2019,Lombart2021}, meaning that the evaluation of the flux at the interface $x_{j-1/2}$ depends on the evaluation of $g_j$ in all bins. As a consequence, the fragmentation flux is a continuous function of mass across interfaces, i.e. $F_{\mathrm{frag}}[g] = F_{\mathrm{frag}}[g](x_{j-1/2},\tau)$. This is in contrast to usual DG solvers in which the numerical flux is discontinuous and must be reconstructed at the interfaces (e.g \citealt{Cockburn1989,Zhang2010}). 

\begin{figure}
\includegraphics[width=\columnwidth,trim=10 290 10 300, clip]{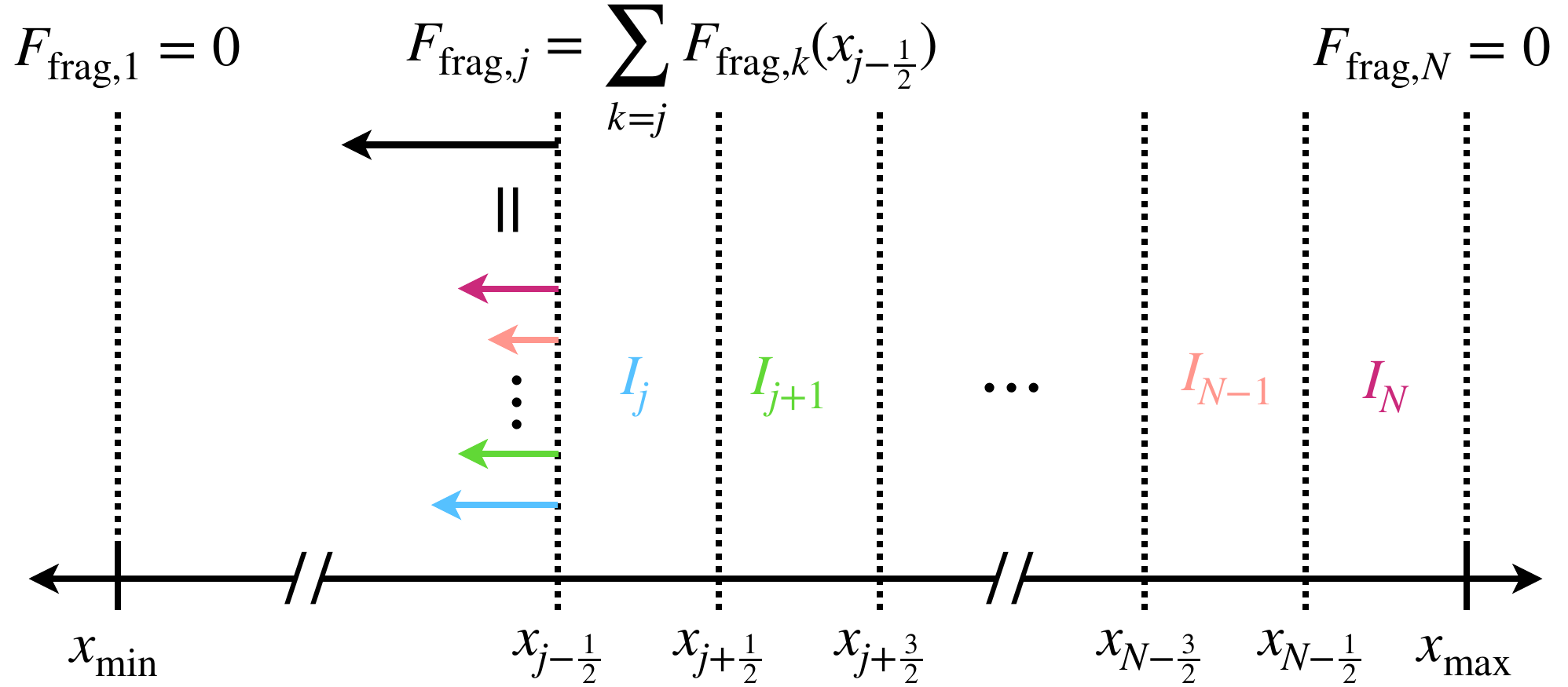}
\caption{Illustration of the fragmentation flux of mass for bin $j$. $F_{\mathrm{frag},j}$ can be seen as the cumulative contribution of sub-fluxes for each $I_k$ transferring mass into the mass range $[x_{\mathrm{xmin}},x_{j-1/2}]$.}
\label{fig:scheme_frag_flux}
\end{figure}

Physically, $F_{\mathrm{frag}}[g](x,\tau)$ represents the cumulative flux from all fragmentation processes which produce particles with mass lower than $x$ from a particle that was greater than $x$, regardless of the bin in which they originated. In practice, since fragmentation tends to produce a mass flux towards smaller masses, we define the flux on the left edge of each bin (i.e. $F_{\mathrm{frag},j}  \equiv F_{\mathrm{frag}}[g](x_{j-1/2})$, see Fig.~\ref{fig:scheme_frag_flux}).  In Fig.~\ref{fig:scheme_frag_flux}, $F_{\mathrm{frag},j}$ is equivalent to the sum of sub-fluxes defined as
\begin{equation}
\begin{aligned}
& \tilde{F}_{\mathrm{frag},k} \\
& \equiv \underbrace{\int\limits_{I_k} \int\limits_{x_{j-1/2}}^{x_{\mathrm{max}}} \frac{\mathcal{K}(u,v)}{uv} g_l'(u,\tau) g_l(v,\tau)}_{\rm A} \underbrace{\int\limits_{x_\mathrm{min}}^{x_{j-1/2}} w \tilde{b}(w,u;v)}_{\rm B}  \mathrm{d}w \mathrm{d}u \mathrm{d}v. 
\end{aligned}
\end{equation}
Term B is the total mass density of particles in the range $[x_\mathrm{xmin},x_{j-1/2}]$ resulting from fragmentation of a particle $u \in [x_{j-1/2},x_\mathrm{xmax}]$. Term A accounts for the rate of fragmentation due to collisions between particles of mass $u$ and a particle of mass $v \in I_k$. Therefore, each term $\tilde{F}_{\mathrm{frag},k}$ transfers mass into $[x_\mathrm{xmin},x_{j-1/2}]$ and contributes to the flux $F_{\mathrm{frag},j} $. The benefit of this convention is that we can ensure mass conservation by setting the outgoing and incoming flux to 0 at $x_{\rm min}$ and $x_{\rm max}$, respectively.
To avoid the prohibitive computational costs of approximating the flux numerically, we take advantage of the polynomial expansion and calculate integrals analytically. For this method to work, both the collision kernel and the distribution of fragments have to be integrable functions, which is the case in this study. This approach maintains a high accuracy and a reasonable computational cost without multiplying the number of sampling points. 

\subsubsection{Procedure to evaluate the flux}
The numerical flux is analytically integrated by using $g_j$, the approximation of $g$ in the bin $I_j$  \citep[see Eq.14 in][]{Lombart2021}. We assume that the distribution of fragments and the collision kernel are explicitly integrable and separable as $\tilde{b}(w,u;v) =  \tilde{b}_3(w)\tilde{b}_1(u)\tilde{b}_2(v)$ and  $\mathcal{K}(u,v)=\mathcal{K}_1(u)\mathcal{K}_2(v)$, which is true for binary fragmentation when using the constant or multiplicative collision kernel:
\begin{equation}
\begin{cases}
& \tilde{b}(w,u;v) = \frac{2}{u} \Rightarrow \tilde{b}_3(w)\equiv 1,\; \tilde{b}_1(u)\equiv \frac{2}{u},\; \tilde{b}_2(v) \equiv 1, \\
& \mathcal{K}_{\mathrm{const}}(u,v) = 1 \Rightarrow \mathcal{K}_1(u) \equiv 1, \; \mathcal{K}_2(v) \equiv 1,\\
& \mathcal{K}_{\mathrm{mul}}(u,v) = uv \Rightarrow \mathcal{K}_1(u) \equiv u, \; \mathcal{K}_2(v) \equiv v.\\
\end{cases}
\end{equation}
The scaled polynomials $p_j(x,\tau)$ replaces $g_j$ to ensure positivity \citep{Liu2019,Lombart2021}
\begin{equation}
\begin{aligned}
&p_j(x,\tau) = \gamma_j (g_j(x,\tau) - \overline{g}_j) + \overline{g}_j,\\
&\gamma_j = \mathrm{min} \left\{ 1,\left| \frac{\overline{g}_j}{m_j-\overline{g}_j} \right| \right\},
\end{aligned}
\end{equation}
where $m_j \equiv \underset{x \in I_j}{\mathrm{min}} g_j(x,\tau)$ and $\overline{g}_j$ refers to the bin average of $g_j$ in $I_j$
\begin{equation}
\overline{g}_j \equiv \frac{1}{h_j} \int_{I_j}g_j(x,\tau) \mathrm{d}x = g_j^0(\tau)
\end{equation}
Therefore, the numerical flux can be written as
\begin{equation}
	\begin{aligned}
		& F_{\mathrm{frag}} (x_{j-1/2},\tau) \\
		& = \sum_{l=1}^N \int_{I_{l}}  \sum_{l'=j}^N \int_{I_{l'}} \sum_{q=1}^{j-1} \int_{I_q} \\
		& \qquad  \frac{w}{uv} \tilde{b}_3(w) \tilde{b}_1(u) \tilde{b}_2(v)  \mathcal{K}_1(u) \mathcal{K}_2(v)  p_{l'}(u,\tau)  p_{l}(v,\tau) \mathrm{d}w \mathrm{d}u \mathrm{d}v,\\
		& =   \sum_{l=1}^N \sum_{l'=j}^N  \sum_{q=1}^{j-1} \sum_{i'=0}^k \sum_{i=0}^k  \gamma_l g_{l}^{i}(\tau) \gamma_{l'} g_{l'}^{i'}(\tau) T_{i',i}(j,l,l',i',i) T_w(j,q) \\
		&  +  \sum_{l=1}^N \sum_{l'=j}^N  \sum_{q=1}^{j-1} \sum_{i'=0}^k (1- \gamma_l )g_{l}^0(\tau) \gamma_{l'} g_{l'}^{i'}(\tau) T_{i',0}(j,l,l',i') T_w(j,q) \\
		&  +  \sum_{l=1}^N \sum_{l'=j}^N  \sum_{q=1}^{j-1} \sum_{i=0}^k  \gamma_l g_{l}^{i}(\tau) (1-\gamma_{l'}) g_{l'}^0(\tau) T_{0,i}(j,l,l',i) T_w(j,q) \\
		&  +  \sum_{l=1}^N \sum_{l'=j}^N  \sum_{q=1}^{j-1} (1- \gamma_l) g_{l}^0(\tau) (1-\gamma_{l'}) g_{l'}^0(\tau) T_{0,0}(j,l,l') T_w(j,q) \\
	\end{aligned}
	\label{eq:frag_flux_num}
\end{equation}
where $k$ is the order of polynomials $p_j$ and
\begin{equation}
\begin{aligned}
	&T_{i',i}(j,l,l',i',i) \equiv \\
	& \qquad \int_{I_l} \int_{I_{l'}}  \frac{\tilde{b}_1(u)}{u} \mathcal{K}_1(u) \phi_{i'}(\xi_{l'}(u)) \frac{\tilde{b}_2(v)}{v} \mathcal{K}_2(v) \phi_{i}(\xi_{l}(v))  \mathrm{d}u \mathrm{d}v, \\
	&T_{i',0}(j,l,l',i') \equiv \\
	& \qquad \int_{I_l} \int_{I_{l'}}  \frac{\tilde{b}_1(u)}{u} \mathcal{K}_1(u) \phi_{i'}(\xi_{l'}(u)) \frac{\tilde{b}_2(v)}{v} \mathcal{K}_2(v)   \mathrm{d}u \mathrm{d}v, \\
	&T_{0,i}(j,l,l',i) \equiv \\
	& \qquad \int_{I_l} \int_{I_{l'}}  \frac{\tilde{b}_1(u)}{u} \mathcal{K}_1(u) \frac{\tilde{b}_2(v)}{v} \mathcal{K}_2(v) \phi_{i}(\xi_{l}(v))  \mathrm{d}u \mathrm{d}v, \\
	&T_{0,0}(j,l,l') \equiv \\
	& \qquad \int_{I_l} \int_{I_{l'}}  \frac{\tilde{b}_1(u)}{u} \mathcal{K}_1(u)  \frac{\tilde{b}_2(v)}{v} \mathcal{K}_2(v)   \mathrm{d}u \mathrm{d}v, \\
	&T_w(j,q) \equiv \int_{I_q} w\tilde{b}_3(w) \mathrm{d}w.
\end{aligned}
\label{eq:flux_T}
\end{equation}
Note that the terms $T_{i',i},\,T_{i',0},\,T_{0,i}$ and  $T_{0,0}$ are calculated using the variables $\xi_{l'}$ and $\xi_{l}$ for numerical stability, and that the negative sign for the fragmentation flux is taken into account in the operator $\bm{L}$ \citep[see Eqs. 16 and 27 in][]{Lombart2021}. All of the analytic formulae are derived and tested in \textsc{Mathematica} before being translated to \texttt{Fortran}. The algorithm is written in \texttt{Fortran} and tested against the \textsc{Mathematica} version for accuracy.

$T_{i','}$, $T_{i',0}$, $T_{0,i}$ and $T_w$ only need to be calculated once at the beginning of the scheme, since they are independent of time. In practice, we combine and store these values in arrays as follows. First we compute four arrays with element $T_{i',i}(j,l,l',i',i) T_w(j,q)$, $T_{i',0}(j,l,l',i') T_w(j,q)$, $T_{0,i}(j,l,l',i) T_w(q)$ and $T_{0,0}(j,l,l') T_w(j,q)$ such that when the subarrays for index j are multiplied respectively by  $\gamma_l g_{l}^{i}(\tau) \gamma_{l'} g_{l'}^{i'}(\tau)$, $(1- \gamma_l )g_{l}^0(\tau) \gamma_{l'} g_{l'}^{i'}(\tau)$, $ \gamma_l g_{l}^{i}(\tau) (1-\gamma_{l'}) g_{l'}^0(\tau)$, $(1- \gamma_l) g_{l}^0(\tau) (1-\gamma_{l'}) g_{l'}^0(\tau) $ and summed over all elements, we obtain the four term of the right-hang side of Eq.~\ref{eq:frag_flux_num}. Finally, the process is repeated for all $j$ to obtain $F_{\mathrm{frag}}$.

\subsection{Integral of the flux}
\label{sec:intflux}
By defining $\mathcal{F}_{\mathrm{frag}}$ as the integral of the numerical flux Eq.~\ref{eq:DG} and \citep[see Eq.12 in][]{Lombart2021}, then together with Eq.~\ref{eq:frag_flux} we obtain
\begin{equation}
\begin{aligned}
&\mathcal{F}_{\mathrm{frag}} (j,k',\tau) = \\
&\int_{I_j} \int\limits_{x_{\mathrm{min}}}^{x_{\mathrm{max}}} \int\limits_x^{x_{\mathrm{max}}} \int\limits_{x_{\mathrm{min}}}^x \frac{w}{uv} \tilde{b}(w,u;v) \mathcal{K}(u,v) g(v,\tau) g(u,\tau) \\
& \qquad \qquad \qquad \qquad \qquad  \times \partial_x \phi_{k'}(\xi_j(x))  \mathrm{d}w \mathrm{d}u \mathrm{d}v \mathrm{d}x,
\end{aligned}
\end{equation}
where $k' \in [\![0,k]\!]$. By splitting the integral over $v$ into two integrals, one for $u \in [x,x_{j+1/2}]$ and the second for $u \in [x_{j+1/2},x_{\mathrm{max}}]$, and by approximating $g$ with the scale polynomials $p$, we obtain
\begin{equation}
	\begin{aligned}
		& \mathcal{F}_{\mathrm{frag}}(j,k',\tau) \\
		& = \sum_{l=0}^N  \sum_{i'=0}^k \sum_{i=0}^k \gamma_j g_j^{i'}(\tau) \gamma_l g_l^i(\tau) \mathcal{T}_{w,u,i'}^1(j,k,i',x_{\mathrm{min}}) \mathcal{T}_{v,i}(j,l,i)  \\
		& + \sum_{l=0}^N  \sum_{i'=0}^k  \gamma_j g_j^{i'}(\tau) (1-\gamma_l) g_l^0(\tau) \mathcal{T}_{w,u,i'}^1(j,k,i',x_{\mathrm{min}}) \mathcal{T}_{v,0}(j,l, \\
		&  + \sum_{l=0}^N  \sum_{i=0}^k (1-\gamma_j) g_j^0(\tau) \gamma_l g_l^i(\tau) \mathcal{T}_{w,u,0}^1(j,k,x_{\mathrm{min}}) \mathcal{T}_{v,i}(j,l,i) \\
		&  +\sum_{l=0}^N  (1-\gamma_j) g_j^0(\tau) (1-\gamma_l) g_l^0(\tau) \mathcal{T}_{w,u,0}^1(j,k,x_{\mathrm{min}}) \mathcal{T}_{v,0}(j,l)\\
		& + \sum_{l=0}^N \sum_{l'=j+1}^N \sum_{i'=0}^k \sum_{i=0}^k \gamma_{l'} g_{l'}^{i'}(\tau) \gamma_l g_l^i(\tau) \\
		& \qquad \qquad \qquad \qquad \qquad \times \mathcal{T}_{w,u,i'}^2(j,k,l',i',x_{\mathrm{min}}) \mathcal{T}_{v,i}(j,l,i) \\
		& + \sum_{l=0}^N \sum_{l'=j+1}^N \sum_{i'=0}^k  \gamma_{l'} g_{l'}^{i'}(\tau) (1-\gamma_l) g_l^0(\tau) \\
		& \qquad \qquad \qquad \qquad \qquad \times \mathcal{T}_{w,u,i'}^2(j,k,l',i',x_{\mathrm{min}}) \mathcal{T}_{v,0}(j,l) \\
		& + \sum_{l=0}^N \sum_{l'=j+1}^N  \sum_{i=0}^k (1-\gamma_{l'}) g_{l'}^0(\tau) \gamma_l g_l^i(\tau) \\
		& \qquad \qquad \qquad \qquad \qquad \times \mathcal{T}_{w,u,0}^2(j,k,l',x_{\mathrm{min}}) \mathcal{T}_{v,i}(j,l,i) \\
		& + \sum_{l=0}^N \sum_{l'=j+1}^N  (1-\gamma_{l'}) g_{l'}^0(\tau) (1-\gamma_l)g_l^0(\tau) \\
		& \qquad \qquad \qquad \qquad \qquad \times \mathcal{T}_{w,u,0}^2(j,k,l',x_{\mathrm{min}}) \mathcal{T}_{v,0}(j,l),
	\end{aligned}
	\label{eq:intflux_frag_num}
\end{equation} 
where
\begin{equation}
\begin{aligned}
&  \mathcal{T}_{w,u,i'}^1(j,k,i',x_{\mathrm{min}}) \equiv \\
&\int_{I_j} \int\limits_{x_{\mathrm{min}}}^x \int\limits_x^{x_{j+1/2}} w \tilde{b}_3(w) \frac{\tilde{b}_1(u)}{u} \mathcal{K}_1(u) \phi_{i'}(\xi_{j}(u)) \\
& \qquad \qquad \qquad \qquad \qquad \times \partial_x \phi_{k'}(\xi_j(x))  \mathrm{d}u \mathrm{d}w \mathrm{d}x, \\
& \mathcal{T}_{w,u,0}^1(j,k,x_{\mathrm{min}}) \equiv  \\
& \int_{I_j} \int\limits_{x_{\mathrm{min}}}^x \int\limits_x^{x_{j+1/2}} w \tilde{b}_3(w) \frac{\tilde{b}_1(u)}{u} \mathcal{K}_1(u) \partial_x \phi_{k'}(\xi_j(x))  \mathrm{d}u \mathrm{d}w \mathrm{d}x, \\
&  \mathcal{T}_{w,u,i'}^2(j,k,l',i',x_{\mathrm{min}}) \equiv \\
&\int_{I_j} \int\limits_{x_{\mathrm{min}}}^x \int\limits_{I_{l'}} w \tilde{b}_3(w) \frac{\tilde{b}_1(u)}{u} \mathcal{K}_1(u) \phi_{i'}(\xi_{l'}(u)) \\
& \qquad \qquad \qquad \qquad \qquad \times \partial_x \phi_{k'}(\xi_j(x))  \mathrm{d}u \mathrm{d}w \mathrm{d}x, \\
& \mathcal{T}_{w,u,0}^2(j,k,l',x_{\mathrm{min}}) \equiv  \\
& \int_{I_j} \int\limits_{x_{\mathrm{min}}}^x \int\limits_{I_{l'}} w \tilde{b}_3(w) \frac{\tilde{b}_1(u)}{u} \mathcal{K}_1(u) \partial_x \phi_{k'}(\xi_j(x))  \mathrm{d}u \mathrm{d}w \mathrm{d}x, \\
& \mathcal{T}_{v,i}(j,l,i) \equiv \int_{I_l} \frac{\tilde{b}_2(v)}{v} \mathcal{K}_2(v) \phi_{i}(\xi_{l}(v)) \mathrm{d}v, \\
& \mathcal{T}_{v,0}(j,l) \equiv \int_{I_l} \frac{\tilde{b}_2(v)}{v} \mathcal{K}_2(v) \mathrm{d}v.
\end{aligned}
\end{equation}
This allows $\mathcal{F}_{\mathrm{frag}}$ to be evaluated similarly to the numerical flux in Eq.~\ref{eq:frag_flux_num}. After a change of variables into $\xi_l$ and $\xi_j$, triple integrals are derived with \textsc{Mathematica} and translated into \texttt{Fortran}. Similar to the arrays in Eq.~\ref{eq:flux_T}, $\mathcal{T}_{w,u,i'}^1$, $\mathcal{T}_{w,u,0}^1$, $\mathcal{T}_{w,u,i'}^2$, $\mathcal{T}_{w,u,0}^2$, $\mathcal{T}_{v,i}$ and  $\mathcal{T}_{v,0}$  need only to be computed once at the beginning of the scheme. Then, $\mathcal{F}_{\mathrm{frag}}$ is reduced to the sum of four terms in Eq.~\ref{eq:intflux_frag_num}. The first term is obtained by computing the product of the subarray of $\mathcal{T}^1_{w,u,i'}(j,k,i',x_{\mathrm{min}}) \mathcal{T}_{v,i}(j,l,i)$ with $\gamma_j g_j^{i'}(\tau) \gamma_l g_l^i(\tau)$ and summing over all elements. The same process is applied for the other terms. Importantly, accuracy depends only on the order of polynomials to approximate $g$, since the integral terms in $F_{\mathrm{frag}}$ and $\mathcal{F}_{\mathrm{frag}}$ are evaluated analytically.

\subsection{CFL condition}
\label{sec:CFL}
Analytic solutions shown in Sect.~\ref{sec:analytic} are of the same form as the linear fragmentation equation \citep{Kostoglou2000,Kumar2014,Liu2019}. Therefore, the CFL condition (CFL) is determined by using the conservative form of the linear fragmentation equation with the flux $F_{\mathrm{linfrag}}$. The DG scheme for order 0 applied to the linear fragmentation equation corresponds to the forward Euler discretisation, i.e.
\begin{equation}
g_j^{0,n+1} = g_j^{0,n} -\frac{\Delta \tau}{\Delta x_j} \left[ F_{\mathrm{linfrag}}(x_{j+1/2},\tau)- F_{\mathrm{linfrag}}(x_{j-1/2},\tau) \right],
\end{equation}
for the $n$-th time step. The stability condition of the scheme is determined by the CFL condition, which is evaluated to ensure the positivity of the bin average $g_j^{0,n+1} = \overline{g}_j^{n+1} >0$ at time-step $n+1$ \citep{Filbet2004,Kumar2007,Lombart2021}, i.e.
\begin{equation}
\begin{aligned}
& g_j^{0,n+1} > 0 \\
& \Rightarrow \Delta \tau \; \underset{j}{\mathrm{sup}} \left( \frac{1}{\Delta x_j} \int_{I_j} \frac{a(v)}{v} \mathrm{d}v \int_{x_{\mathrm{min}}}^{x_{j-1/2}} u b(u,x_j) \mathrm{d}u \right) <1,
\end{aligned}
\end{equation}
where $b(u,x_j)=2/x_j$, $a(v)=v$ for the multiplicative kernel and $a(v)=1$ for the constant kernel. Details of the derivation are given in Appendix~\ref{appendix:CFL}. Unfortunately, we did not succeed in obtaining a better CFL criterion with the method based on the Laplace transform presented in \citet{Laibe2022}.
\begin{table}
	\centering
	\caption{CFL condition for linear and quadratic fragmentation rates.}
	\label{table:CFL}
	\sisetup{
		detect-all,
		table-format			= 1.1e-1,
		scientific-notation		= true,
		table-auto-round		= true,
		round-integer-to-decimal	= true,
		}
	\begin{tabular*}{1.0\columnwidth}{ l c S } \toprule
                %%%%
                	Type		&	$\mathcal{K}(x,y)$		&	{$\Delta \tau_{\rm CFL}$}	\\\midrule
                %%%%
		constant		&	$1$			&	2.8		\\
		multiplicative		&	$xy$		&	4e-3	\\\bottomrule
		%%%%
	\end{tabular*}
\end{table}
Table~\ref{table:CFL} gives the CFL condition for the constant and multiplicative kernels. For the constant collision kernel case, it is required to apply the change of time variable detailed in Eq.~\ref{eq:time_var}. The DG method writes 
\begin{equation}
\frac{\mathrm{d} \bm{g}_j\left(\tau' \right)}{\mathrm{d} \tau'}  = \exp(-\tau') \bm{L}[g],
\end{equation}
where $\bm g$ and $\bm L$ are defined in Eqs. 14 and 16 in \citet{Lombart2021}. Therefore, the time step changes into 
\begin{equation}
\Delta \tau' \rightarrow \exp(-\tau'_n) \Delta \tau'.
\label{eq:timestep_kconst}
\end{equation}
In practice, we use the same CFL than the multiplicative kernel to reduce numerical errors.

%-----------------------------------------------------------------------------------------------------------------
\section{Numerical results}{}
\label{sec:num}

The numerical scheme presented in Sect.~\ref{sec:dg} is tested against the exact solutions detailed in Sect.~\ref{sec:analytic}.  Tests are performed with a limited number of bin, $N=20$, consistent with constrains from 3D hydrodynamical simulations. The algorithm flowchart in detail is given in the github repository in Sect.~\ref{data_github}.

\subsection{Evaluation of errors}
\label{sec:errors}
The strategy for error measurements is similar to \citet{Lombart2021}. The experimental order of convergence (EOC, \citealt{Liu2019}) and the efficiency of the algorithm are analysed. The continuous and discrete $L^1$ norm writes
\begin{equation}
\begin{aligned}
   &e_{\mathrm{c},N}(\tau) \equiv \sum_{j=1}^N \frac{h_j}{2} \sum_{\alpha=1}^R \omega_{\alpha} |g_j(x_j^{\alpha},\tau) - g(x_j^{\alpha},\tau)|,\\
   &e_{\mathrm{d},N}(\tau) \equiv \sum_{j=1}^N h_j |g_j(\hat{x}_j,\tau) - g(\hat{x}_j,\tau)|,
 \end{aligned}
   \label{eq:errL1}
\end{equation}
where $g$ and $g_j$ are the analytic and the numerical solutions of the linear fragmentation equation, $R=16$ is the number of points used for the gaussian quadrature method, and $\hat{x}_j\equiv \sqrt{x_{j-1/2}x_{j+1/2}}$ is the geometric mean of bin $I_j$. 

The EOC is defined as
\begin{equation}
   \mathrm{EOC} \equiv \frac{\ln\left( \frac{e_N(\tau)}{e_{2N}(\tau)}\right)}{\ln(2)},
   \label{eq:EOC}
\end{equation}
where $e_N$ is the error evaluated for $N$ cells and $e_{2N}$ for $2N$ cells. For the calculation of the EOC, the numerical errors are calculated after one time-step at time $\tau =10^{-3}$ for constant and multiplicative kernels in order to avoid time stepping errors. The total mass density of the system, the first moment of $g(x,\tau)$, writes 
\begin{equation}
      M_{1,N}(\tau) \equiv \int\limits_{x_{\mathrm{min}}}^{x_{\mathrm{max}}} g(x,\tau) \mathrm{d}x=    \sum_{j=1}^N h_j g_j^0(\tau) .
\end{equation}
The absolute error of the total mass density is given by
\begin{equation}
   e_{M_{1,N}}(\tau) \equiv \frac{|M_{1,N}(\tau) - M_1|}{M_1},
\end{equation}
where $M_1$ is the first moment of order of $g$ for the exact solution and is constant in time. Here, we compare only absolute gains in order to interface the algorithm with hydrodynamical solvers.

\subsection{Implementation details}
\label{subsec:benchmark_frag}
Numerical solutions for the constant and multiplicative kernels are benchmarked against the exact solutions given in Eqs.~\ref{eq:sol_kconst}  and \ref{eq:sol_kmul}. Simulations are performed for a mass interval $x \in [10^{-6},10^3]$. Tests are performed with \textsc{Fortran} in double precision and errors are calculated with \textsc{Python}. Numerical solutions are shown only for polynomials of order $k=0,1,2,3$. For $k>3$,  arithmetics of large numbers generate non negligible errors. The initial condition \citep[see Eq.17 in][]{Lombart2021} is evaluated by a Gauss-Legendre quadrature method with five points to satisfy the analytical solution at time $\tau=0$, with the majority of the mass contained in large grains. After a finite time $\tau$, fragmentation has redistributed the mass to smaller grain sizes and we can compare the analytical and numerical solutions for accuracy. Simulations are performed with constant time steps of $\Delta \tau_{\rm lin} = 10^{-3} $ for constant and multiplicative kernels to satisfy the respective CFL conditions (see Table~\ref{table:CFL}).  Simulations are performed sequentially on an Intel Core i$7$ $2.8\mathrm{GHz}$. We use the \texttt{gfortran v11.2.0} compiler.

\subsection{Multiplicative kernel}
\label{subsec:kmul_tests}

%combine figures
\begin{figure*}
\centering
\includegraphics[width=0.9\textwidth]{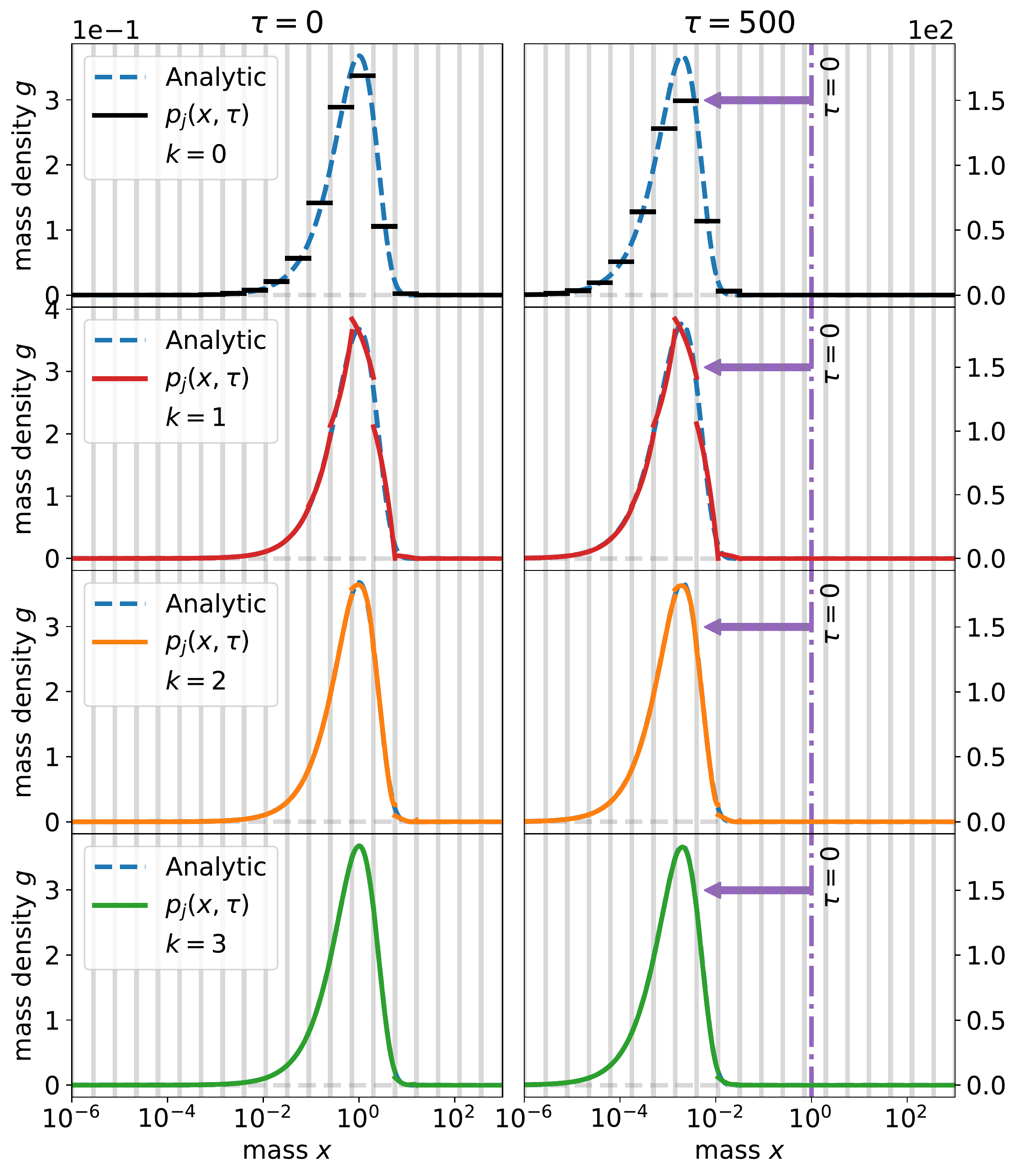}
\caption{Test for the multiplicative kernel. Solid lines show the numerical solution $p_j(x,\tau)$ for $N=20$ bins and $k = 0,1,2,3$ at time  $\tau=0$ (left) and $\tau=500$ (right).  The exact solution $g(x,\tau)$ is given by the blue dashed line. Vertical grey lines represent the boundaries of the bins while the purple dashed lines in the panels on the right indicate the peak of the initial distribution to highlight the evolution due to fragmentation indicated by the purple arrow (note the nearly three orders of magnitude difference between the left and right vertical axes). The accuracy improves for larger values of $k$. Order $3$ recovers the peak of the mass distribution with an accuracy of order $\sim 0.1 \%$.}
\label{fig:kmul_linlog}
\end{figure*}

\begin{figure*}
\centering
\subfloat[][]{\includegraphics[width=0.9\columnwidth]{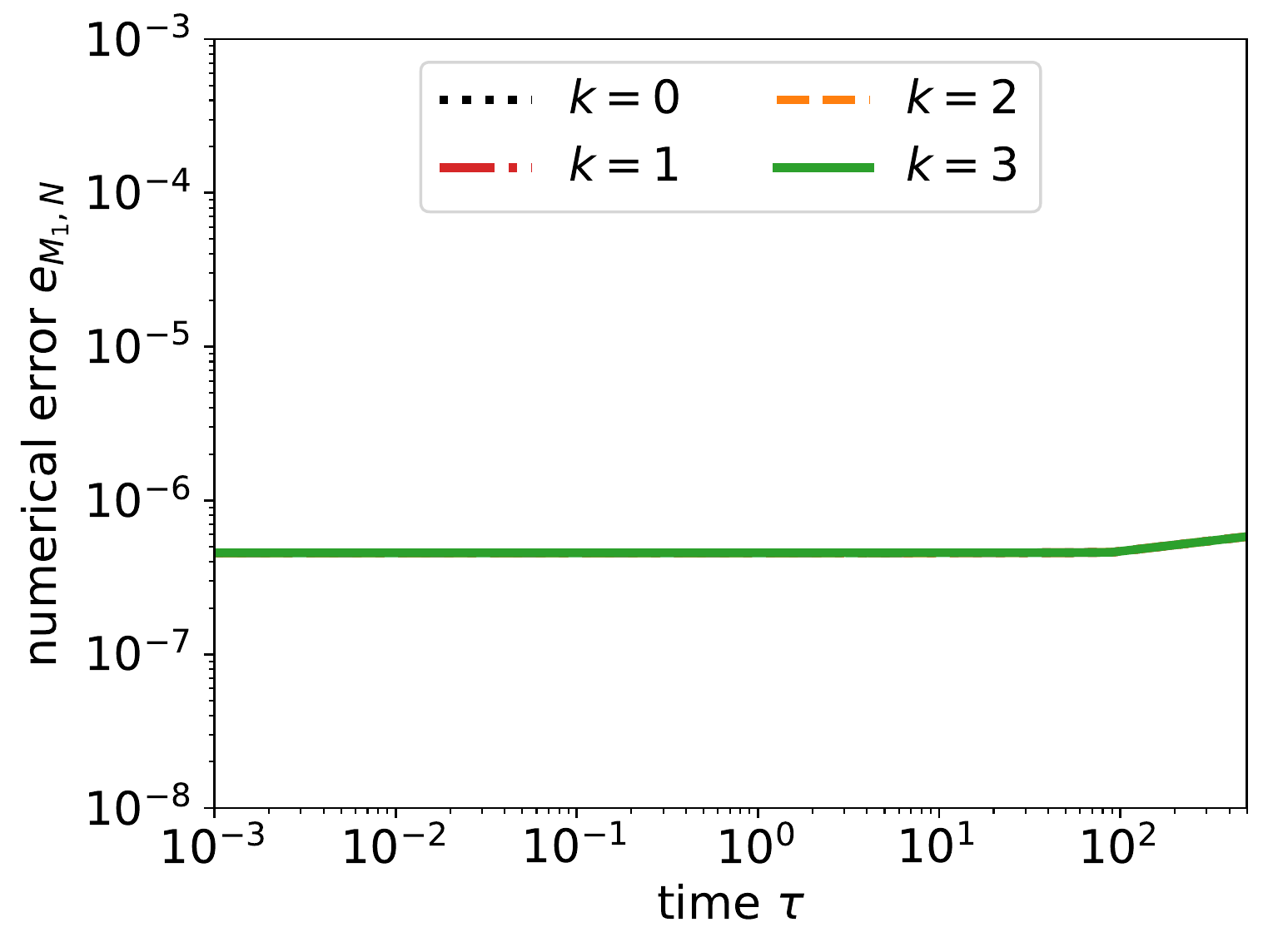}\label{fig:kmul_a}}
\subfloat[][]{\includegraphics[width=0.9\columnwidth]{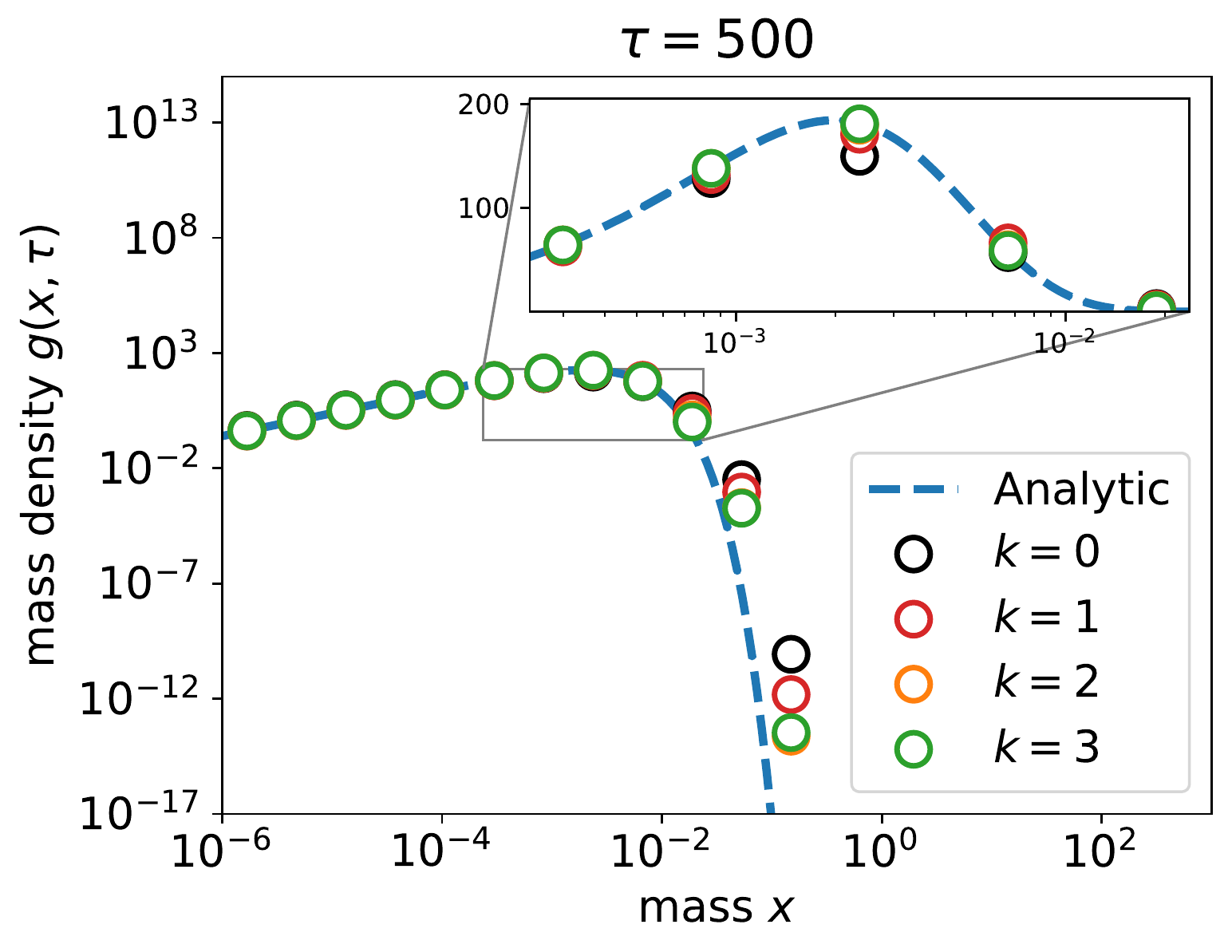}\label{fig:kmul_b}}\\
\subfloat[][]{\includegraphics[width=0.9\columnwidth]{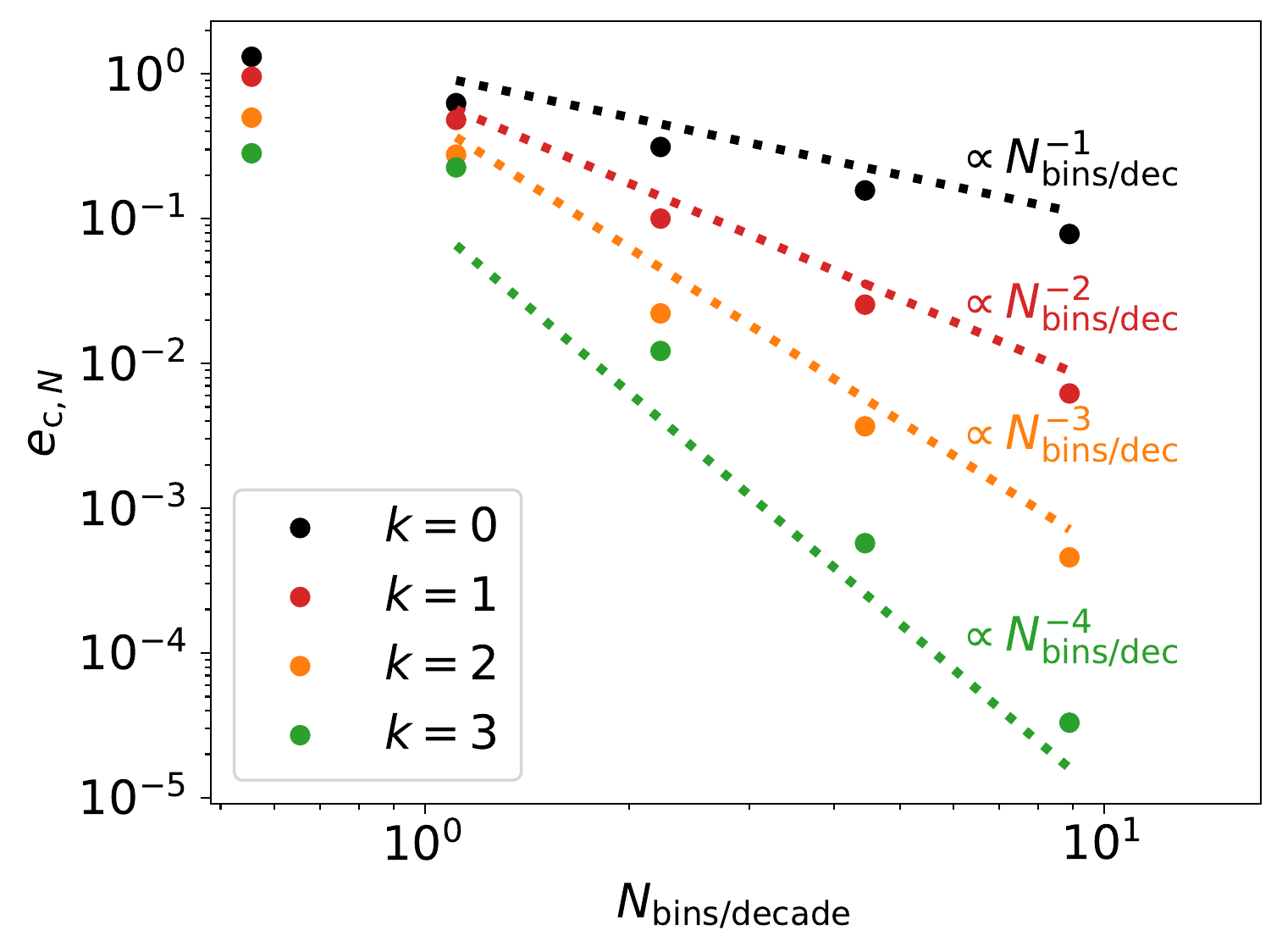}\label{fig:kmul_c}}
\subfloat[][]{\includegraphics[width=0.9\columnwidth]{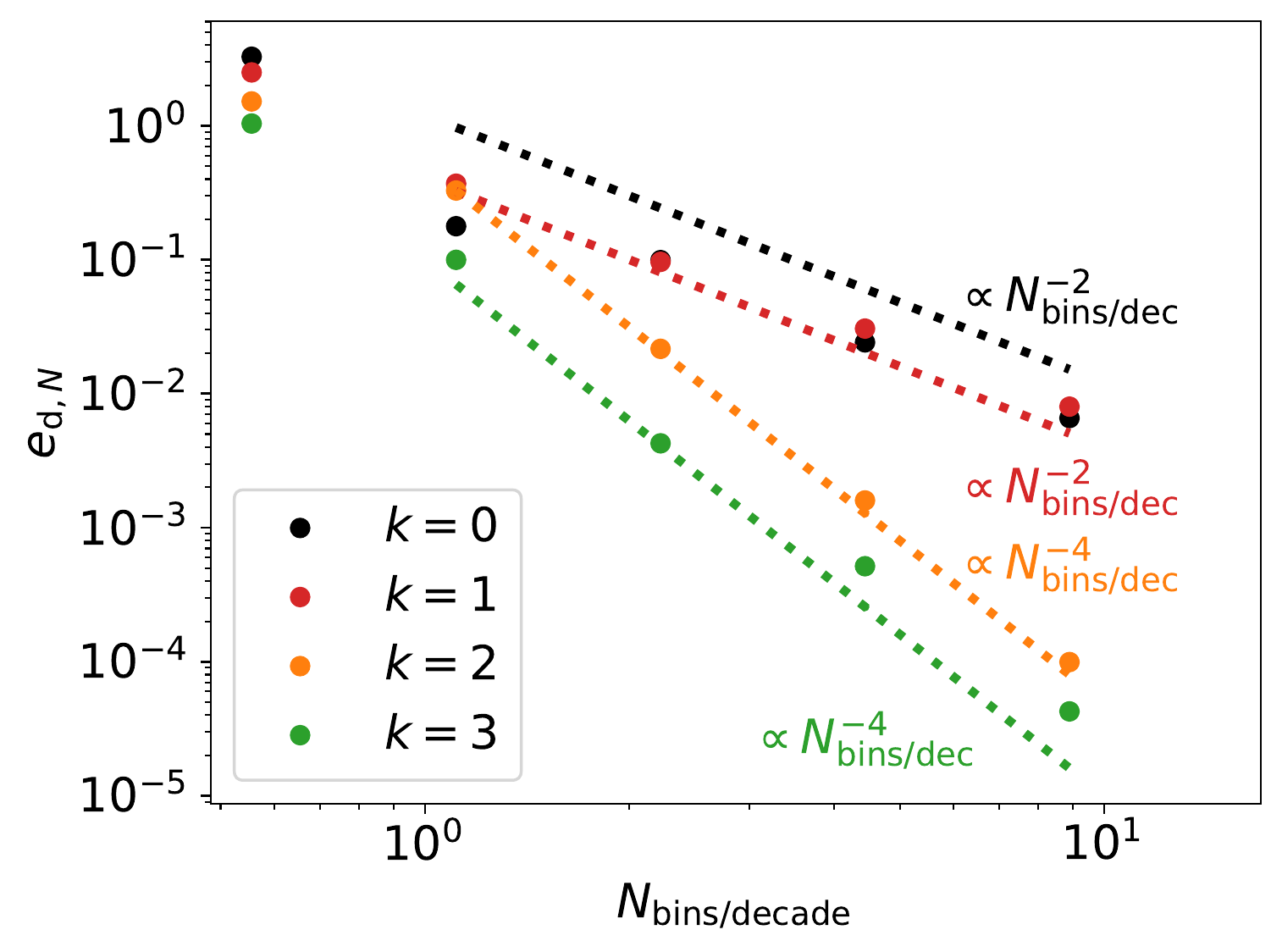}\label{fig:kmul_d}}\\
\subfloat[][]{\includegraphics[width=0.9\columnwidth]{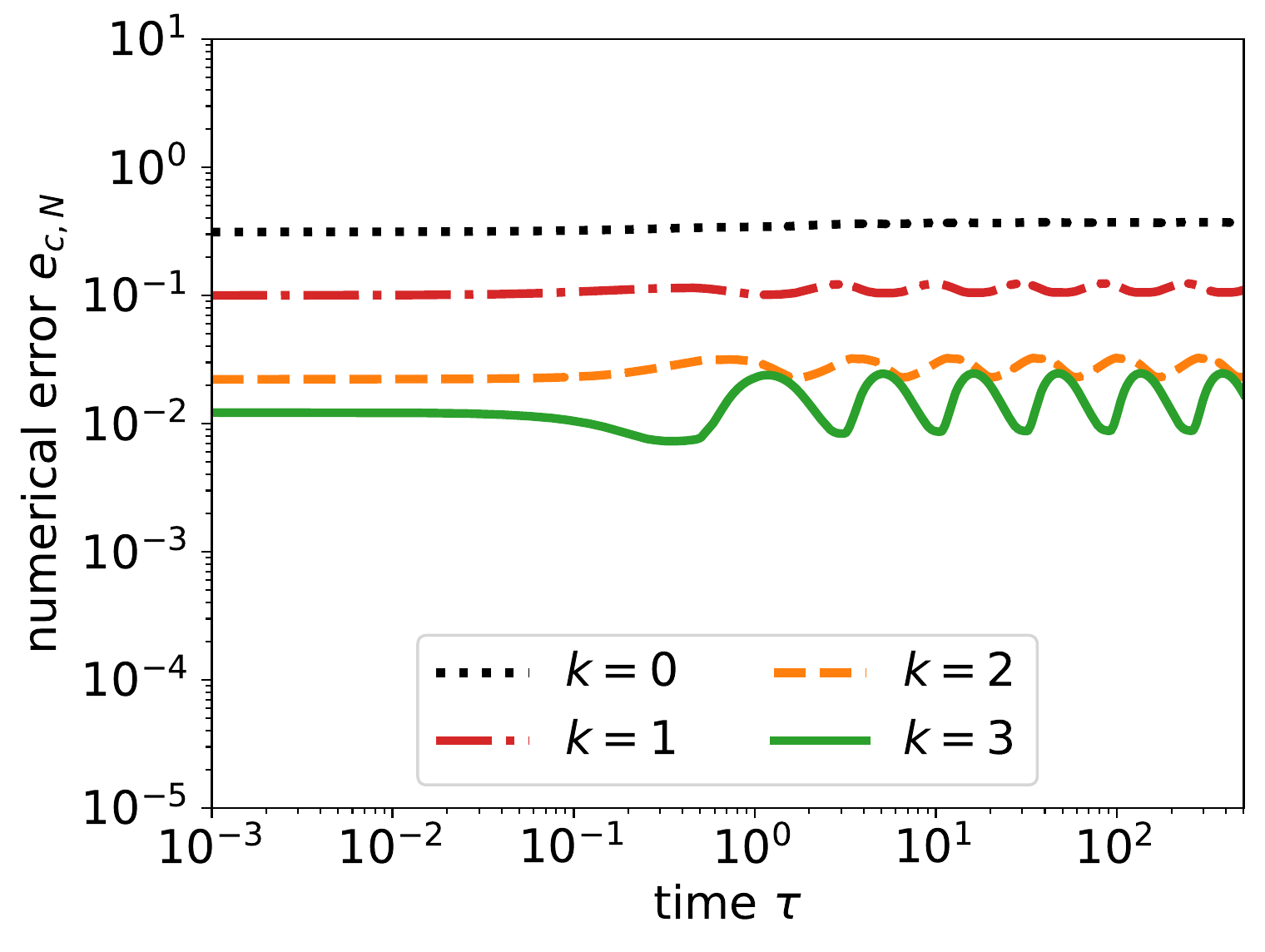}\label{fig:kmul_e}}
\subfloat[][]{\includegraphics[width=0.9\columnwidth]{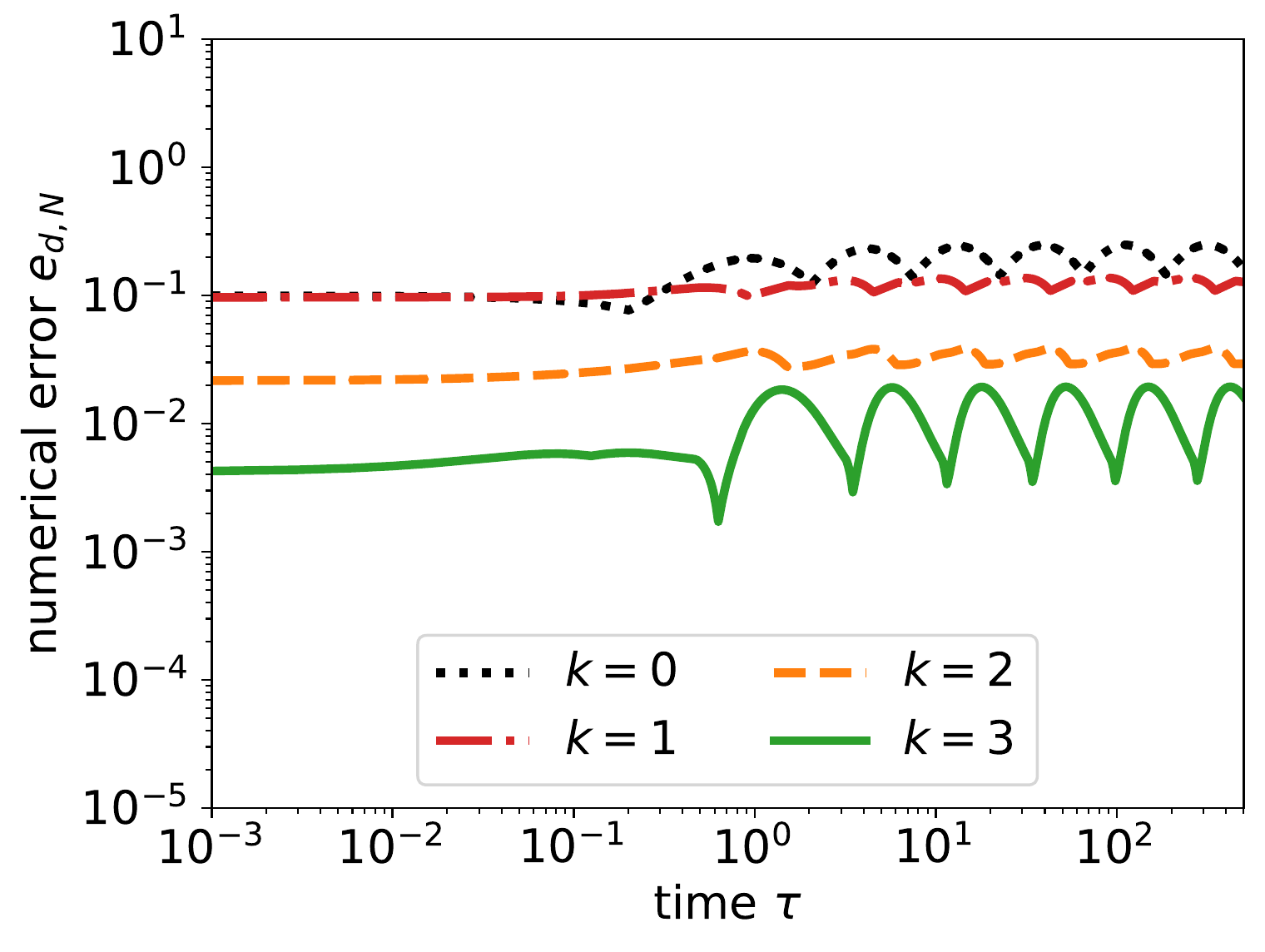}\label{fig:kmul_f}}

\caption{Test for the multiplicative kernel with $N=20$ bins. Fig.~\protect\subref{fig:kmul_a}: evolution of the absolute error $e_{M_1,N}$ on the moment $M_{1,N}$ for $N=20$ bins. The mass is conserved. Fig.~\protect\subref{fig:kmul_b}: numerical solution $p_j(x,\tau$) evaluated with the geometric mean $\hat{x}_j$ over each bin $I_j$. A zoom-in of the peak of the distribution shows that an absolute error of $\sim 0.1 - 1\%$ is reached with $k = 1,2,3$, compared to $20\%$ obtained with $k = 0$. Accuracy in the exponential tail is improved by a factor $1000$ with $k = 3 $ compared to $k = 0$. Figs.~\protect\subref{fig:kmul_c},\protect\subref{fig:kmul_d}: the continuous $L^1$ error $e_{c,N}$ and the discrete $L^1$ error $e_{d,N}$ are plotted versus the number of bins per decade. The experimental order of convergence is EOC$=k+1$ for $e_{c,N}$ and for $e_{d,N}$ EOC$=k+1$ for polynomials of odd orders and  EOC$=k+2$ for polynomials of even orders. An accuracy of $0.1\%$ is achieved with more than $10$ bins/decade for $k=0,1$, with $\sim 6$ bins/decade for $k = 2$ and with $\sim 4$ bins/decade for $k = 3$. An accuracy of $1\%$ is achieved with $\sim 10$ bins/decade for $k = 0,1$, with $\sim 3$ bins/decade for $k = 2$ and $\sim 2$ bins/decade for $k = 3$. Figs.~\protect\subref{fig:kmul_e},\protect\subref{fig:kmul_f}: time evolution of the $L^1$ continuous and discrete norms, showing they remain bounded at large times.}
\label{fig:kmul}

\end{figure*}

\subsubsection{Conservation of mass and positivity of numerical solutions}
\label{subsubsec:kmul_positivity}
Numerical solutions with orders of polynomials varying from $k=0$ to $k=3$ for $N=20$ bins are benchmarked against analytical solutions at $\tau=500$, in Fig.~\ref{fig:kmul_linlog}. The numerical solutions are positive due to the combination of the slope limiter and the Strong Stability Preserving Runge-Kutta (SSPRK) time stepping \citep[see Sect. 3.4 and Sect. 3.5 in][]{Lombart2021}. Fig.~\ref{fig:kmul},\subref{fig:kmul_a} shows the absolute error $e_{M_{1,N}}$ for $N=20$ bins from $\tau=10^{-3}$ to $\tau =500$. As expected, the mass is conserved for all orders $k$.

\subsubsection{Accuracy of the numerical solution}
\label{subsubsec:accuracy}
The accuracy of the numerical solutions improves with order of the polynomials. Fig.~\ref{fig:kmul},\subref{fig:kmul_b} shows the numerical solution obtained at $\tau=500$ in log-log scale. The numerical diffusion is drastically reduced  (by up to a factor 1000) in the exponential tail as the order of the scheme increases. The major part of the total mass is localised around the maximum of the curve. The zoomed-in panel shows a linear scaling of this region, where it is evident that numerical solutions with order $k = 1,2,3$ achieve absolute errors of order $\sim 0.1 - 1\%$ while errors of order $\sim 20\%$ are obtained with $k = 0$. 

\subsubsection{Convergence of the numerical scheme}
\label{subsubsec:convergence}
Figs.~\ref{fig:kmul},\subref{fig:kmul_c},\subref{fig:kmul_d} show the numerical errors introduced in Sect.~\ref{sec:errors} at $\tau=10^{-3}$. We use $e_{\mathrm{c},N}$ and $e_{\mathrm{d},N}$ from order 0 to order 3 as a function of the number of bins per decade $N_{\mathrm{bin}/\mathrm{dec}}$ to determine the EOC independently from the mass range. On the left, the EOC  for the continuous $L^1$ norm is of order $k+1$.  On the right, the EOC for the discrete $L^1$ norm  is of order $k+2$ for odd polynomials, and $k+1$ for even polynomials. From $e_{d,N}$, the expected accuracy of order $\sim 0.1\%$ is achieved with more than $10$ bins/decade with $k=0,1$, $\sim 6$ bins/decade with $k=2$ and $\sim 4$ bins/decade with $k=3$. Errors of order $\sim 1\%$ is reached with $\sim 10$ bins/decade for $k=0,1$, with $\sim 3$ bins/decade for $k=2$, and with $\sim 2$ bins/decade for $k=3$.

\subsubsection{Stability of the numerical scheme}
\label{subsubsec:stability_time}
Figs.~\ref{fig:kmul},\subref{fig:kmul_e},\subref{fig:kmul_f} show the time evolution for $e_{c,N}$ and $e_{d,N}$. Despite only using $N = 20$ bins, both the continuous and discrete errors for each order remain bounded over the entire time interval (albeit with different accuracy).

\subsection{Constant kernel}
\label{subsec:kconst_tests}

%combine figures
\begin{figure*}
\centering
\includegraphics[width=0.9\textwidth]{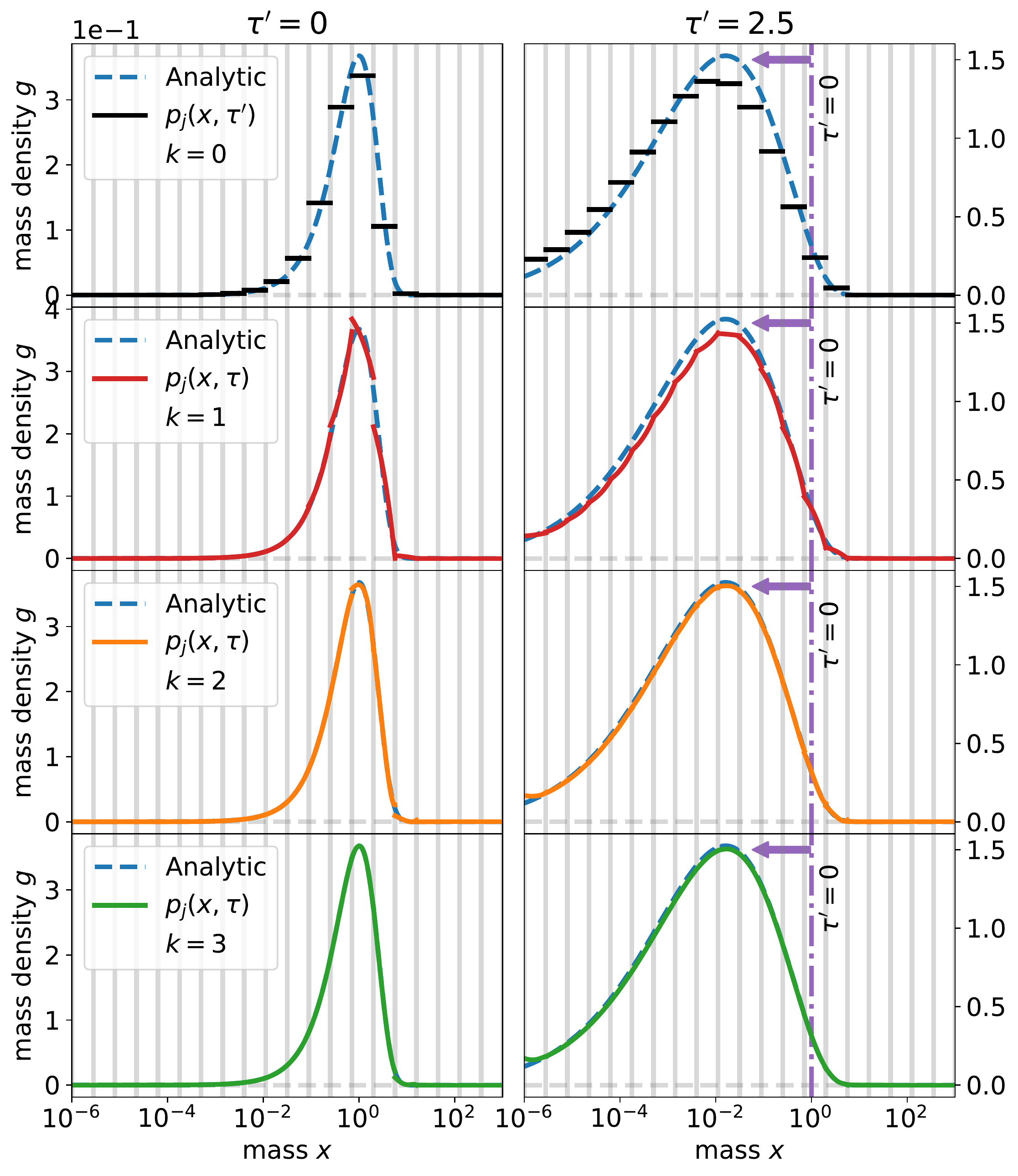}
\caption{As in Fig.~\ref{fig:kmul_linlog}, but using the constant kernel from $\tau'=0$ to $\tau'=2.5$.}
\label{fig:kconst_linlog}
\end{figure*}

\begin{figure*}
\centering

\subfloat[][]{\includegraphics[width=0.9\columnwidth]{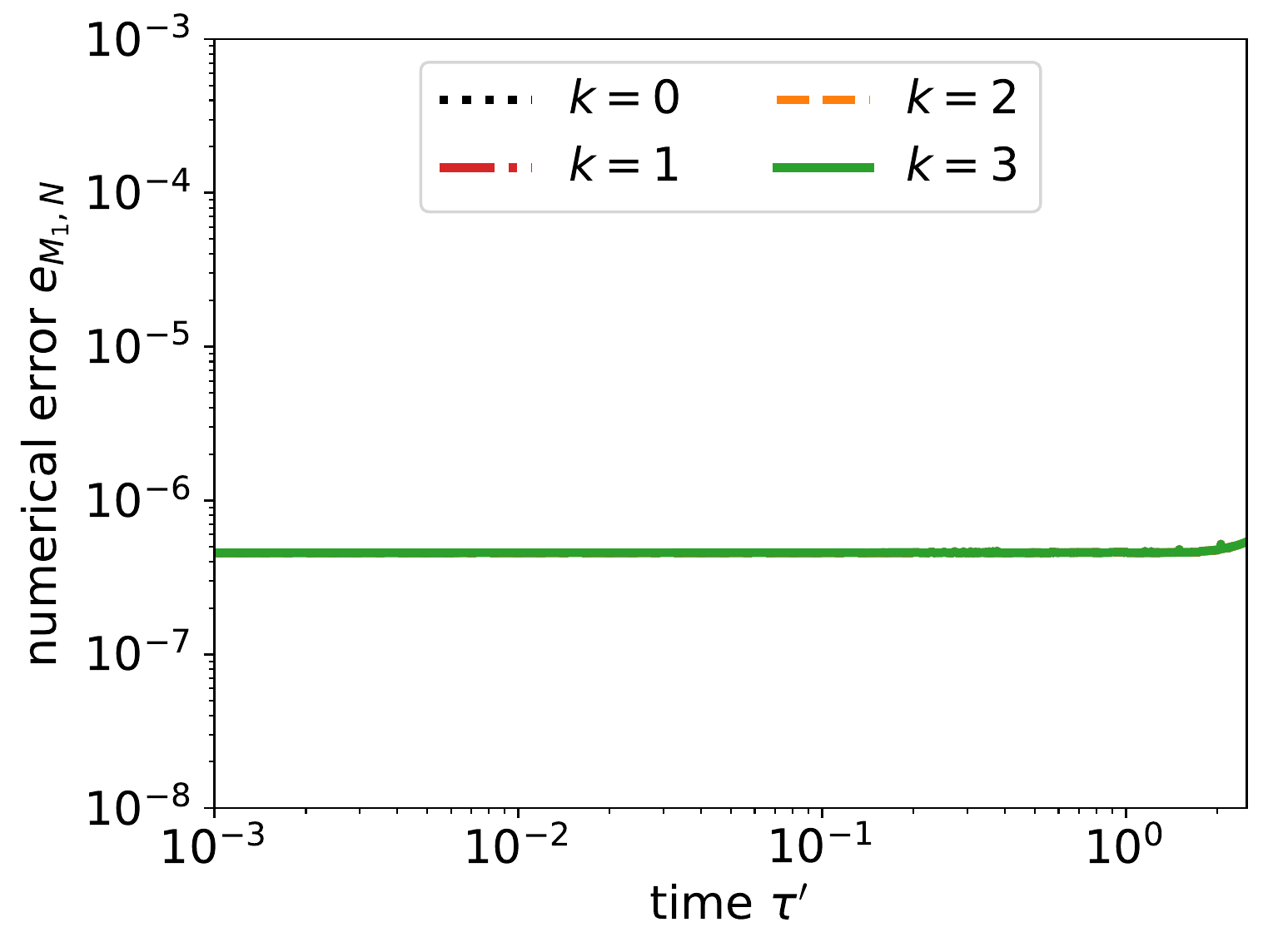}\label{fig:kconst_a}}
\subfloat[][]{\includegraphics[width=0.9\columnwidth]{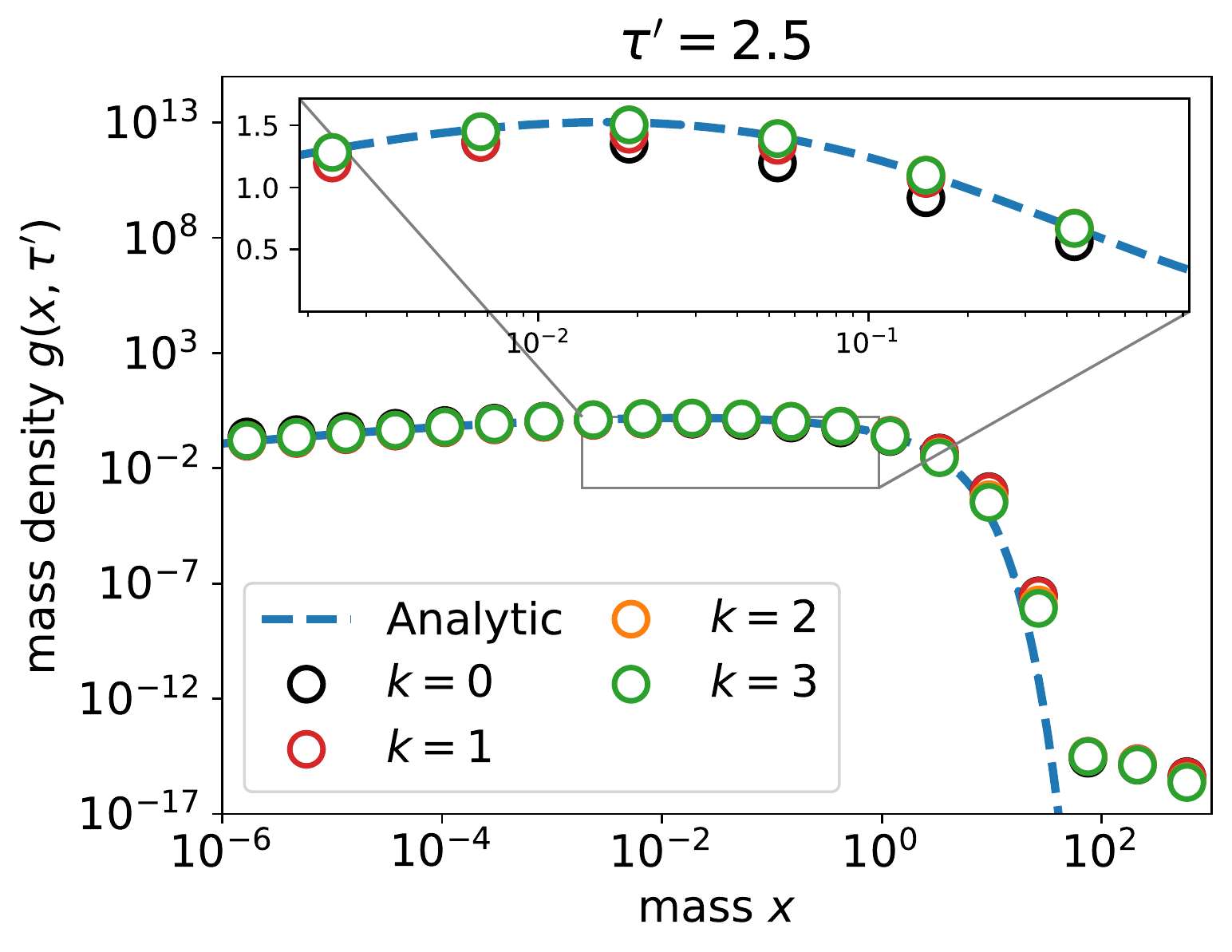}\label{fig:kconst_b}} \\
\subfloat[][]{\includegraphics[width=0.9\columnwidth]{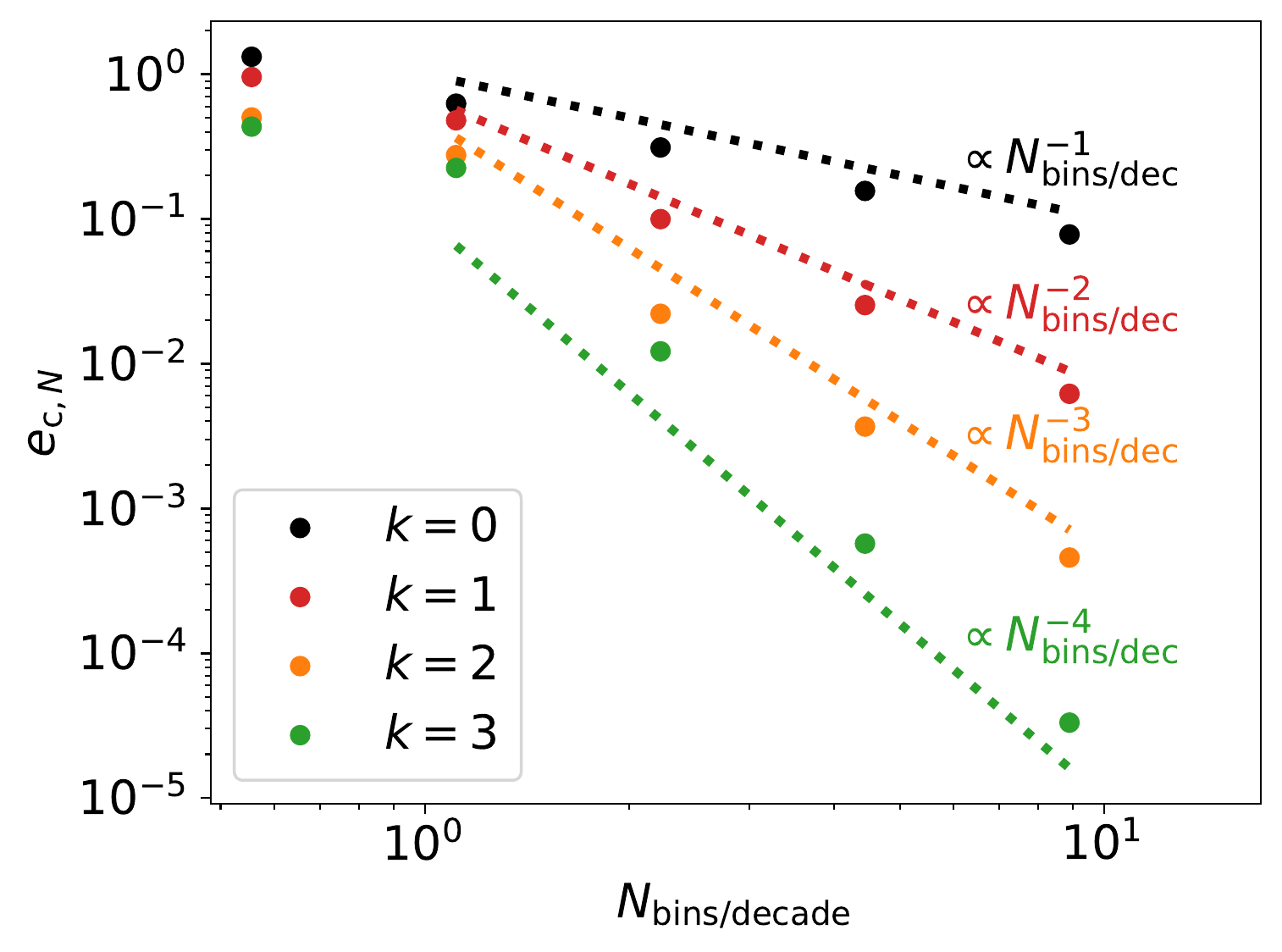}\label{fig:kconst_c}}
\subfloat[][]{\includegraphics[width=0.9\columnwidth]{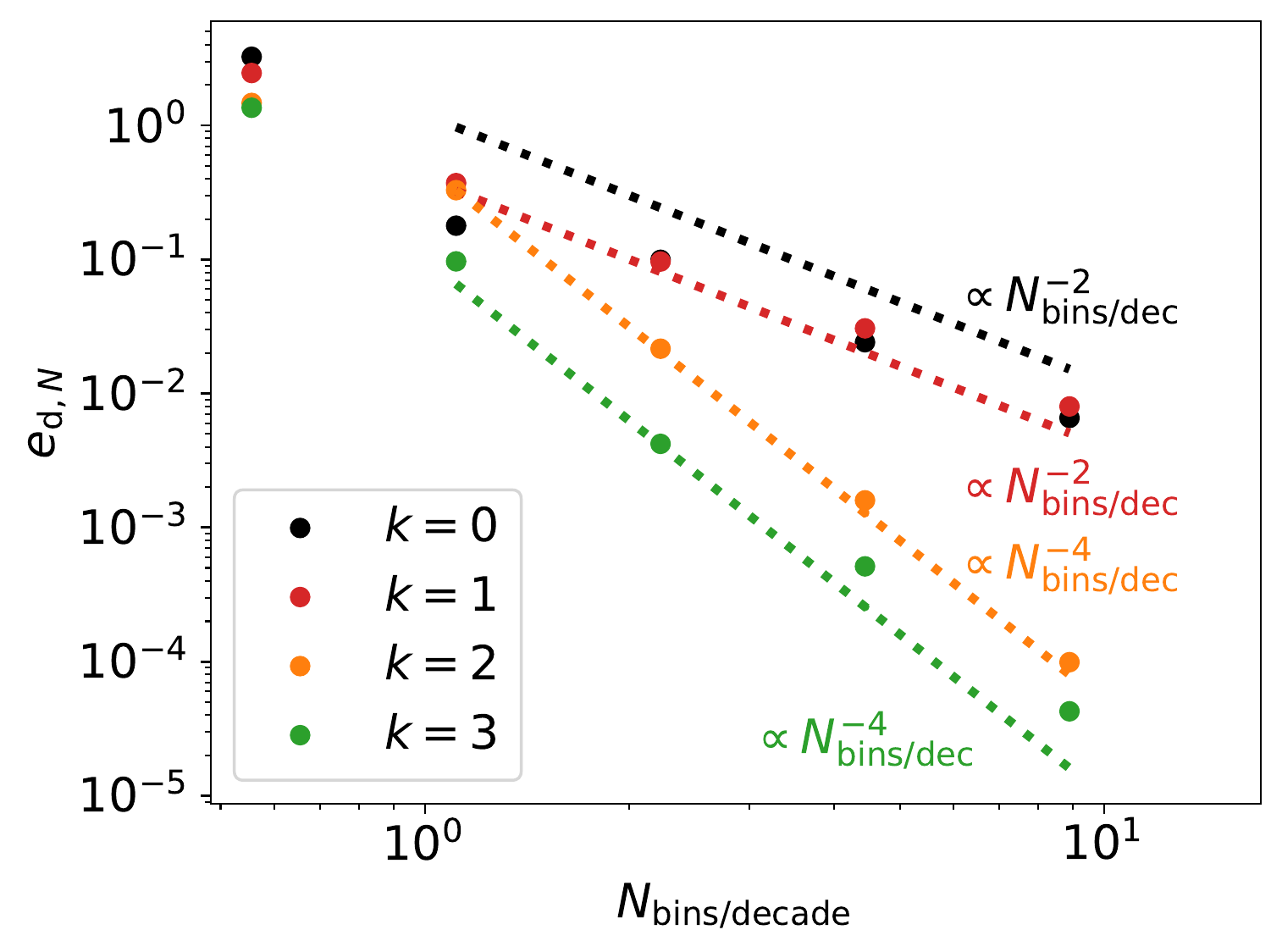}\label{fig:kconst_d}}\\
\subfloat[][]{\includegraphics[width=0.9\columnwidth]{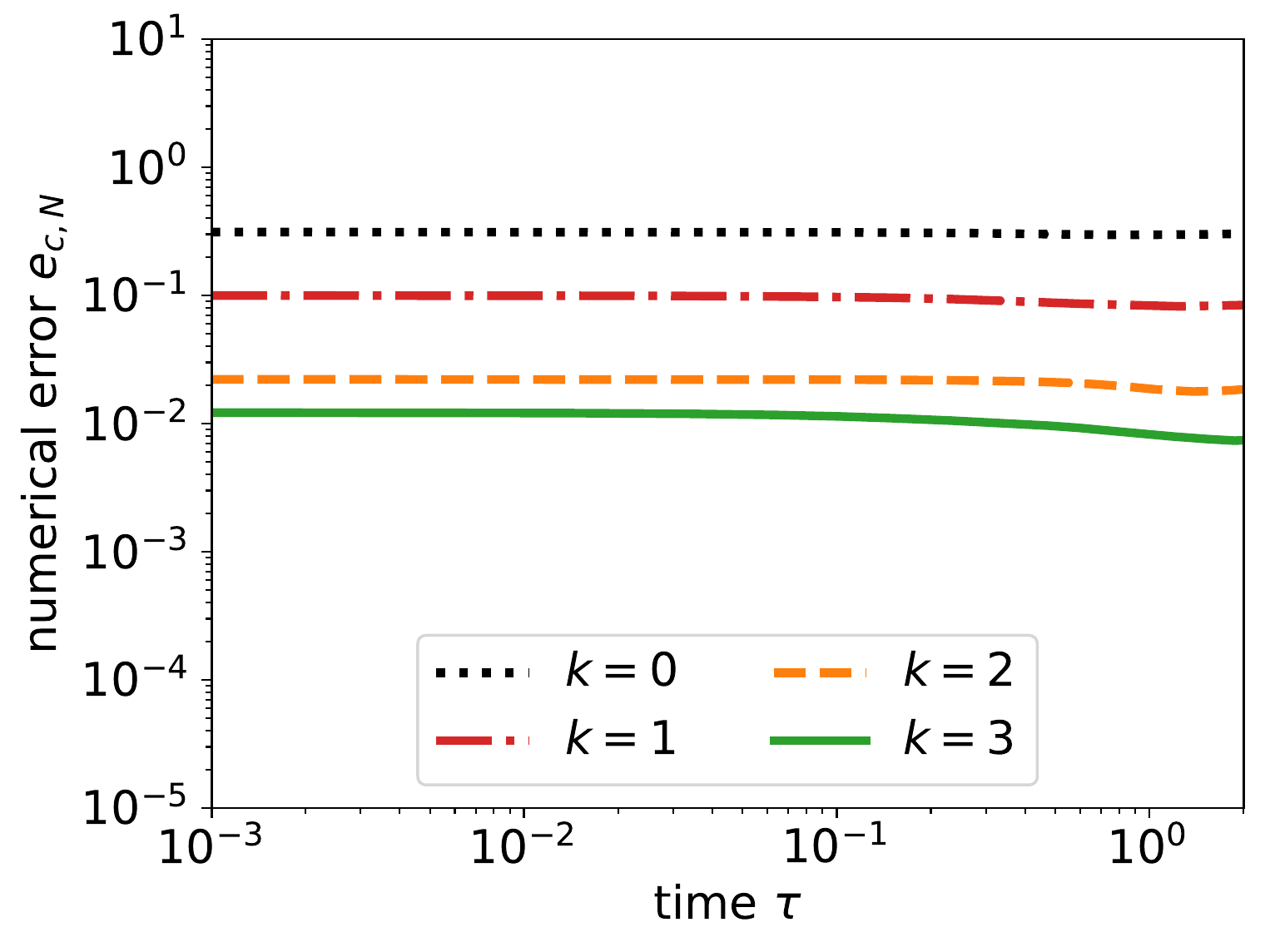}\label{fig:kconst_e}}
\subfloat[][]{\includegraphics[width=0.9\columnwidth]{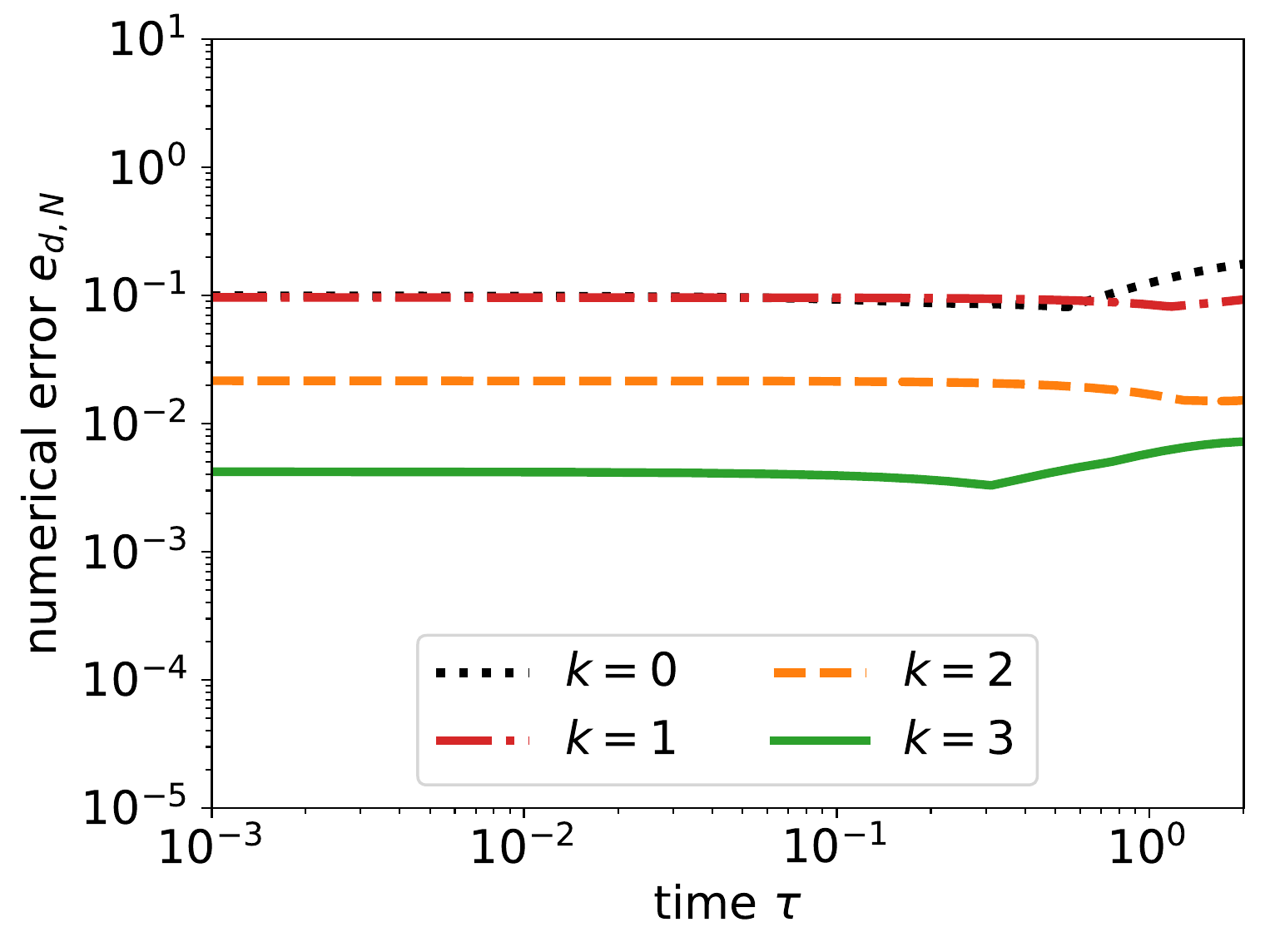}\label{fig:kconst_f}}

\caption{As in Fig.~\ref{fig:kmul} but using the constant kernel.}

\label{fig:kconst}

\end{figure*}

For constant case, the analytic solution Eq.~\ref{eq:sol_kconst} is evaluated for 1000 points logarithmically spaced with the function \texttt{integrate} in library \texttt{scipy}. Figures \ref{fig:kconst_linlog} and \ref{fig:kconst} show the test results for the constant kernel (comparable to Figs.\ref{fig:kmul_linlog} and \ref{fig:kmul}). Numerical solutions preserve positivity and errors have similar magnitude compares to the multiplicative case after the same number of time steps.

%-----------------------------------------------------------------------------------------------------------------
\section{Discussion}
\label{sec:discussions}
The high-order numerical scheme based on the Discontinuous Galerkin method presented in Sect.~\ref{sec:dg} requires arithmetic of large numbers for polynomials of high-order. Thus, in its current form, order $k=3$ is the maximum limit of the algorithm. Table~\ref{tab:time} gives the elapsed wall time for a single time-step for each order when using the constant or multiplicative kernel. 
\begin{table}
\begin{center}
\begin{tabular}{ccccc}
\hline
  & $k=0$ & $k=1$ & $k=2$ & $k=3$ \\ 
\hline
time (s) & $9.10^{-4}$ & $2.10^{-2}$ & $5.10^{-2}$ & $9.10^{-2}$ \\  
\end{tabular}
\caption{Elapsed wall time in second for a single time-step for each orders when using constant or multiplicative kernel.}
\label{tab:time}
\end{center}
\end{table}
Using the code \texttt{PHANTOM} \citep{Price2018} with a typical multifluid simulation with 10 dust sizes embedded in a protoplanetary disc with $10^6$ SPH particles and $32$ cpu, one hydrodynamical time-step takes $\sim 1\,\mathrm{s}$. The CFL condition detailed in Sect.\ref{sec:CFL} for fragmentation will likely impose a sub-cycling of the hydrodynamic time-step, potentially creating a computational bottleneck. To avoid this bottleneck, two strategies may help to reach a one-to-one ratio: i) using a specific change of variable to reduce the number of time-steps \citep{Carrillo2004,Goudon2013}, and ii) using GPU parallelisation, since the fragmentation solver can be used independently for each SPH particle, which is highly parallelisable. Moreover, implicit time-stepping can be implemented in a manageable way because the number of dust bins has been kept minimal. We will be testing these strategies to further gain accuracy and computation efficiency in the future.

The collisional fragmentation model presented in this study considers that only one of the two colliding particles breaks up. No rearrangement of mass is allowed, i.e. mass transfer during the fragmentation process is not taken into account \citep{Kostoglou2000,Banasiak2019}. Actually, the form of Eq.~\ref{eq:frag_cont_DL} implicitly assumes the mass of fragments in a collision does not exceed the mass of the parent particle $y$ (see Eq.~\ref{eq:mass_cons} and surrounding text). However, experiments have shown that some fragmentation events do lead to mass transfer between colliding grains, including cases where one of the colliders ends up with more mass than it started \citep{Bukhari_Syed2017,Blum2018}. To account for this more general case, we could consider the impact as a two-step process. The first step is the same as before with only one of the impacting grains fragmenting. The second step occurs immediately afterwards with one of the fragments sticking to the unbroken grain, leading to an increase in mass. While the latter is technically a coagulation process, it differs from the typical events modelled in the coagulation equation because interactions with the unbroken grain are limited to the distribution of fragments of the broken grain (as opposed to the entire grain size distribution). In such a scenario, Eq.~\ref{eq:frag_cont_DL} has to be modified as follows \citep{Banasiak2019}
\begin{equation}
\begin{aligned}
\frac{\partial f (x,\tau)}{\partial \tau} = & \frac{1}{2} \int\limits_{0}^{\infty} \int\limits_{0}^{\infty} \mathcal{K}(y,z) \overline{b}(x,y;z) f(y,\tau) f(z,\tau) \mathrm{d}y \mathrm{d}z \\
& - f(x,\tau) \int\limits_{0}^{\infty} \mathcal{K}(x,y) f(y,\tau) \mathrm{d}y,
\end{aligned}
\label{eq:newfrag_cont_DL}
\end{equation}
where $\overline{b}(x,y;z)$ is the distribution of fragments and a symmetric function in $y$ and $z$. The symmetric property of function $\mathcal{K}$ and $\overline{b}(x,y;z)$ explains the factor $1/2$ (i.e. collisions are counted twice in the integral). The new criterion for mass conservation is
\begin{equation}
\int\limits_0^{y+z} x \overline{b}(x,y;z) \mathrm{d}x = y+z, \; \forall x>y+z,\, \overline{b}(x,y;z) =0.
\label{eq:new_mass_cons}
\end{equation}
Eq.~\ref{eq:frag_cont_DL} can be recovered from Eq.~\ref{eq:newfrag_cont_DL} by considering the symmetric property of the distribution of fragments \citep{Banasiak2019}
\begin{equation}
\overline{b}(x,y,;z) = \tilde{b}(x,y;z)H(y-x) + \tilde{b}(x,z;y)H(z-x),
\end{equation}
where $H$ is the Heaviside function. In future work, the DG scheme will be applied to the general non-linear fragmentation equation Eq.~\ref{eq:newfrag_cont_DL} in its conservative form. 

Physically, the fragmentation kernel used in astrophysics is the ballistic kernel defined as 
\begin{equation}
   K\! \left(x,y \right) \equiv \left(1-\beta\! \left(x,y, \Delta v\right)\right) \Delta v \!\left(x,y\right) \sigma \! \left(x,y\right),
\end{equation}
where $\Delta v$ is the mean relative velocity between two grains of masses $x$ and $y$, $\sigma$ is the mean effective cross section of collision and $\beta$ denotes the mean sticking probability of the grains. In this study, where only fragmentation is considered, then $\beta =0$. The microphysics of collisions are encoded inside $\sigma$ and $\Delta v$, those parameters depending \textit{a priori} on the sizes of the colliding grains or the kinetic and thermodynamical parameters of the surrounding medium. If we assume the grains are not charged, then we can ignore the complicating effects of electrostatic forces and $\sigma$ reduces to the sum of the radii of the colliding dust grains. Unfortunately, even with these simplifications, the ballistic kernel does not yield analytic solutions and numerical solutions are required. Future developments will be to implement the ballistic kernel in the general non-linear fragmentation model (Eq.~\ref{eq:newfrag_cont_DL}).

By nature, fragmentation and coagulation are stochastic processes, but random fluctuations in the solution can not be computed with Eq.~\ref{eq:frag_cont_DL}. Given that such fluctuations cannot be constrained by current observations, this is not a critical limitation of the model. More importantly, there is nothing in the scheme that would prevent stochasticity from being added to the model. Additional physical processes, such as fragmentation of aggregates, can be implemented in the solver by extending Eq.~\ref{eq:frag_cont_DL}. In this case, the collision kernel $\mathcal{K}$ has to be rewritten to include the adapted cross-section \citep{Friedlander2000}
\begin{equation}
\mathcal{K}(x,y) \sim (x^{1/D_f}+y^{1/D_f}) \Delta v(x,y),
\end{equation}
where $D_f$ is the fractal dimension calculated by the collision algorithm, such as Particle-Cluster Aggregation (PCA) or Cluster-Cluster Aggregation (CCA) \citep{Dominik1997,Paszun2009,Dominik2016}.

%-----------------------------------------------------------------------------------------------------------------
\section{Conclusion}
\label{sec:conclusion}
Fragmentation resulting from colliding grains is an important process that helps to regulate the smaller end of the evolving dust size distribution. Chemical, thermal and dynamical processes strongly depend on the dust size distribution, highlighting the need for accurate dust models that include coagulation and fragmentation. We have presented a high-order DG algorithm that accurately solves the non-linear fragmentation equation on a reduced mass grid of $\sim 20$ bins in order to provide an efficient algorithm that can feasibly couple with 3D hydrodynamic codes. Importantly, the DG scheme meets all of the physical and numerical requirements to accomplish this goal, including: i) a strictly positive mass density maintained over the entire numerical grid made possible by a strong stability preserving Runge Kutta time solver combined with a slope limiter, ii) conservation of mass at machine precision, iii) accuracy of order $0.1-1\%$ is reach by high-order discretisation in time and mass space and iv) a fast algorithm with manageable time and memory costs due to analytically computed integrals and the ability to run with a limited number of bins.

The non-linear fragmentation model presented in this study describes the fragmentation of only one of the two colliding grains. The most general case consists of fragmentation and potential mass transfer between two colliding aggregates of arbitrary size . Although this model has been widely used in astrophysics \citep{Safronov1972,Tanaka1996,Birnstiel2010,Hirashita2021}, numerical algorithms have never been compared to exact solutions, which to our knowledge have only been derived recently \citep{Banasiak2019}. Another crucial ingredient affecting the mass flux in physical models is coagulation. The methods we have developed in Paper I and Paper II (this study) are compatible with those developed earlier by \citet{Lombart2021} for coagulation. We will combine these processes in a future study and test their performance on real astrophysical mixtures.

\section*{Acknowledgements}
This work has received funding from the Ministry of Science and Technology, Taiwan (MOST 110-2636-M-003-001). ML acknowledges funding from the ERC CoG project PODCAST No 864965. This project has received funding from the European Union's Horizon 2020 research and innovation programme under the Marie Sk\l odowska-Curie grant agreement No 823823. MAH acknowledges support from the Excellence Cluster ORIGINS, which is funded by the Deutsche Forschungsgemeinschaft (DFG, German Research Foundation) under Germany Excellence Strategy - EXC-2094 - 390783311 and partial funding by the Deutsche Forschungsgemeinschaft (DFG, German Research Foundation) - 325594231. We used \textsc{Mathematica} \citep{Mathematica}.  We thank the anonymous referee for useful comments and discussions.

\section*{Data availability}
\label{data_github}
The data and supplementary material underlying this article are available in the repository "nonlinear\_frag" on GitHub at \url{https://github.com/mlombart/nonlinear\_frag.git}. Figures can be reproduced following the file \texttt{README.md}. The repository contains data and Python scripts used to generate figures.

%\label{lastpage}
\bibliography{biblio_frag_paper2}

\appendix
\section{Linear fragmentation}
\label{ap:lin_frag}
The linear fragmentation model is a fragmentation process in which the breakup of grains is driven spontaneously by external forces where dust collisions are rare, meaning that the fragmentation is not due to grain-grain collisions \citep{Cheng1990,Kostoglou2000}. Linear fragmentation is an ubiquitous phenomenon in nature that underlies processes such as polymer degradation \citep{Ziff1985}, breakup of liquid droplets \citep{Pruppacher2010,Khain2018}, grinding or crushing of rocks \citep{Austin1971}, industrial crystallisation \citep{Mazzarotta1992}, emulsification \citep{Ramkrishna2000,Hakansson2009} and disruption of grains \citep{Draine1996,Hoang2019}. Similar to the Smoluchowski coagulation equation \citep{Muller1928,Safronov1972}, the linear fragmentation equation is formalised by a deterministic mean-field approach \citep{Redner1990}. With no known generic analytic solutions, a large literature exists concerning the linear fragmentation equation since the 1950s (\citealt{Melzak1957,Srivastava1971,Ziff1985,Redner1990,Kostoglou2007,Kumar2013,Banasiak2019}).

\subsection{Conservative form}
\begin{figure}
\includegraphics[width=\columnwidth,trim=100 150 50 50, clip]{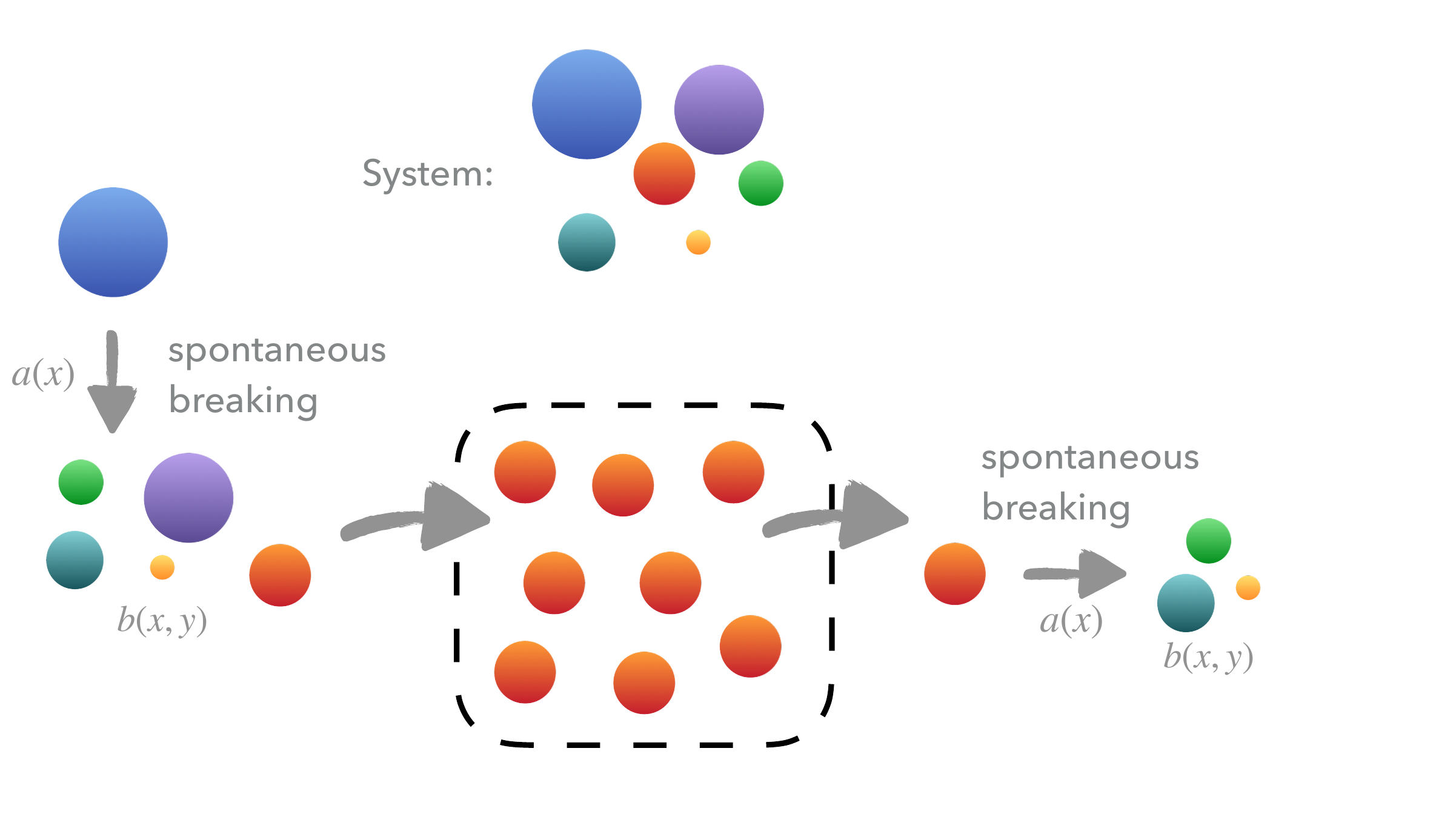}
\caption{Scheme of the spontaneous fragmentation equation Eq.~\ref{eq:linfrag_cont}. The number density of orange grains, mass $m$, increases due to fragmentation of the blue grains with mass greater than $m$, and decreases due to fragmentation of orange grains.}
\label{fig:scheme_linfrag}
\end{figure}

The linear fragmentation equation was formalised in a mean-field approach by \citet{Melzak1957}. The by-products of fragmenting aggregates are called fragments and are assumed to be spherical. Spatial correlations are neglected. For physical systems involving the fragmentation of aggregates made of a large number of monomers, it is convenient to assume a continuous mass distribution. The population density of grains within a differential mass range $\mathrm{d}m$ is characterised by its number density $n(m)$. The continuous linear fragmentation equation is given by
\begin{equation}
\begin{aligned}
\frac{\partial n (m,t)}{\partial t } = - a(m) n(m,t) + \int\limits_m^{\infty} b(m,m') a(m') n(m',t) \mathrm{d}m',
\end{aligned}
\label{eq:linfrag_cont}
\end{equation}
where $t$ denotes time, $n(m,t)$ is the number density function per unit mass for particles of mass $m$. The first term on the right-hand side of Eq.~\ref{eq:linfrag_cont} accounts for the loss of particles of mass $m$ due to their breaking up into smaller particles (breaking of an orange particle on Fig.~\ref{fig:scheme_linfrag}), the coefficient $a(m)$ being the rate of fragmentation for particles of mass $m$. The second term of Eq.~\ref{eq:linfrag_cont} represents the increase of particles of mass $m$ produced by the fragmentation of a particle of mass $m'>m$ (left side on Fig.~\ref{fig:scheme_linfrag}),  the coefficient $b(m,m')$ being the distribution of fragments.

If $n_0(m) = n(m,0)$, is the initial number density distribution per unit mass, then the total mass density, the total number density of grains and the mean mass of the initial distribution can be written as
\begin{equation}
M = \int\limits_0^{\infty} mn_0(m) \mathrm{d}m, \, N_0 = \int\limits_0^{\infty} n_0(m) \mathrm{d}m,\, m_0=\frac{M}{N_0}.
\end{equation}
The linear fragmentation equation is written in dimensionless form by using the following dimensionless variables \citep{Kostoglou2007}:
\begin{equation}
  \left\{ 
  \begin{aligned}
    & x \equiv m/m_0,\,y \equiv m'/m_0,\, \tau = a(m_0) t, \\
    & \tilde{a}(x)=a(m)/a(m_0),\, \tilde{b}(x,y)=m_0 b(m,m'),\\
    & f(x,\tau) = m_0 \, n(m,t)/N_0.
  \end{aligned}
  \right.
\end{equation}
Here, $a(m_0)$ is a normalising constant with dimension $[s^{-1}]$. We use the variables $x$, $\tau$, and $f$ for the dimensionless mass, time, and number density, respectively, to be consistent with the existing literature \citep[e.g.][]{Kostoglou2007,Banasiak2019} such that Eq.~\ref{eq:linfrag_cont} writes
\begin{equation}
\begin{aligned}
\frac{\partial f (x,\tau)}{\partial \tau } = - \tilde{a}(x) f(x,\tau) + \int\limits_x^{\infty} \tilde{b}(x,y) \tilde{a}(y) f(y,\tau) \mathrm{d}y.
\end{aligned}
\label{eq:linfrag_cont_DL}
\end{equation}
 \citet{Kumar2007} has shown that Eq.~\ref{eq:linfrag_cont_DL} can be equivalently written in conservative form, as follows
\begin{equation}
\left\{
   \begin{aligned}
   &\frac{\partial g \left( x,\tau \right) }{\partial \tau} + \frac{\partial F_{\mathrm{frag}} \left[ g \right] \left( x,\tau \right)}{\partial x} = 0,\\
   &F_{\mathrm{frag}} \left[ g \right] \left( x,\tau \right) = - \int\limits_x^{\infty} \int\limits_0^x u \tilde{b}(u,v) \frac{\tilde{a}(v)}{v} g(v,\tau) \mathrm{d}u \mathrm{d}v,
   \end{aligned}
\right.
\label{eq:linfrag_cons_DL}
\end{equation}
where $g(x,\tau) \equiv xf(x,\tau)$ is the mass density of grains per unit mass, and $F_{\mathrm{frag}} [g] (x,\tau)$ is the flux of mass density across the mass $x$ triggered by fragmentation.

\subsection{Discontinuous Galerkin algorithm}
\begin{figure}
\centering
\includegraphics[width=0.9\columnwidth]{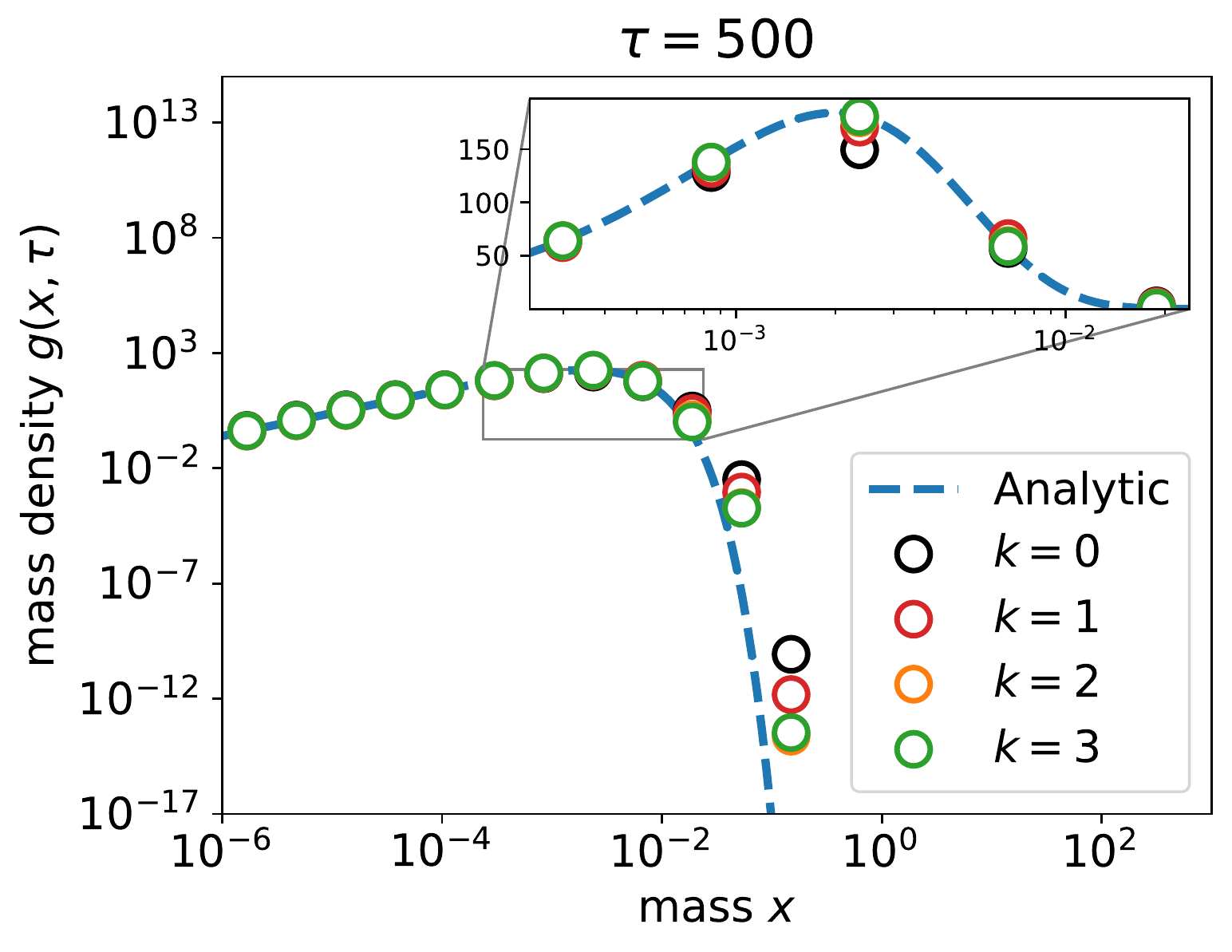}
\caption{Test for the linear fragmentation rate with $N=20$ bins. Numerical solution $p_j(x,\tau$) evaluated with the geometric mean $\hat{x}_j$ over each bin $I_j$. A zoom-in of the peak of the distribution shows that an absolute error of $\sim 0.1 - 1\%$ is reached with $k = 1,2,3$, compared to $20\%$ obtained with $k = 0$. Accuracy in the exponential tail is improved by a factor $1000$ with $k = 3 $ compared to $k = 0$. The missing points have values below $10^{-17}$.}
\label{fig:lin_frag}
\end{figure}

Similar to the non-linear fragmentation equation, the Discontinuous Galerkin method is applied to the linear fragmentation \citep{Liu2019,Lombart2021} and tested against the exact solutions \citep{Ziff1985}. Performances of the DG scheme are similar to the application on the non-linear fragmentation and only a result for the linear fragmentation $a(x)=x$ with 20 bins is shown here in Fig.~\ref{fig:lin_frag} for example.

\section{Constant collision kernel}
\label{ap:kernel_const}
The derivation of Eq.~\ref{eq:kconst_N} is given here:
\begin{equation}
\begin{aligned}
	\frac{\mathrm{d} N}{\mathrm{d}t} &= - N^2 + N \int_0^{\infty} \int_x^{\infty}  \frac{b(x/y)}{y} f(y,t) \mathrm{d}y \mathrm{d}x\\
	& = - N^2 + N \int_0^{\infty} \int_0^{\infty} H(y-x) \frac{b(x/y)}{y} f(y,t) \mathrm{d}y \mathrm{d}x\\ 
	& =  - N^2 + N \int_0^{\infty} \int_0^{\infty} \underbrace{H(y-x)}_{\text{applied on $x$}} \frac{b(x/y)}{y} f(y,t) \mathrm{d}x \mathrm{d}y\\
	& = - N^2 + N \int_0^{\infty} \int_0^y  \frac{b(x/y)}{y} f(y,t) \mathrm{d}x \mathrm{d}y\\
	 & \underbrace{=}_{w\equiv x/y}  - N^2 + N \int_0^{\infty} \int_0^1  b(w) f(y,t) \mathrm{d}w \mathrm{d}y\\
	& = (b_0-1)N^2,
\end{aligned}
\end{equation}
where $H$ is the Heaviside function and $b_0=\int_0^1 b(w) \mathrm{d}w$, with $w=x/y$.

\section{CFL condition}
\label{appendix:CFL}
The conservative form of the linear fragmentation writes \citep{Kumar2014,Liu2019}
\begin{equation}
\left\{
   \begin{aligned}
   &\frac{\partial g \left( x,\tau \right) }{\partial \tau} + \frac{\partial F_{\mathrm{linfrag}} \left[ g \right] \left( x,\tau \right)}{\partial x} = 0,\\
   &F_{\mathrm{linfrag}} \left[ g \right] \left( x,\tau \right) = - \int\limits_x^{\infty} \int\limits_0^x u b(u,v) \frac{a(v)}{v} g(v,\tau) \mathrm{d}u \mathrm{d}v,
   \end{aligned}
\right.
\label{eq:linfrag_cons_DL}
\end{equation}
where $a(v)$ is the rate of fragmentation for particle of mass $v$ and $b(u,v)$ is the distribution of fragments of mass $u$ resulting from a break-up of particle of mass $v$. The DG scheme applied to Eq.\ref{eq:linfrag_cons_DL} for order 0 writes
\begin{equation}
g_j^{0,n+1} = g_j^{0,n} -\frac{\Delta \tau}{\Delta x_j} \left[ F_{\mathrm{linfrag}}(x_{j+1/2},\tau)- F_{\mathrm{linfrag}}(x_{j-1/2},\tau) \right],
\label{eq:EF}
\end{equation}
where
\begin{equation}
\begin{aligned}
&F_{\mathrm{linfrag}}(x_{j+1/2},\tau) = \\
& \qquad \qquad - \sum_{l=j+1}^N \int_{I_l} \int_{x_{\mathrm{min}}}^{x_{j+1/2}} u b(u,v) \frac{a(v)}{v}   g_l^0(\tau) \mathrm{d}u \mathrm{d}v,
\end{aligned}
\end{equation}
which is the numerical representation of the flux \citep{Liu2019}. Therefore, we obtain
\begin{equation}
\begin{aligned}
& F_{\mathrm{linfrag}}(x_{j+1/2},\tau)-F_{\mathrm{linfrag}}(x_{j-1/2},\tau) \\
& =  - \sum_{l=j+1}^N \int_{I_l} \int_{x_{\mathrm{min}}}^{x_{j+1/2}} u b(u,v) \frac{a(v)}{v}   g_l^0(\tau) \mathrm{d}u \mathrm{d}v \\
& \qquad \qquad + \sum_{l=j}^N \int_{I_l} \int_{x_{\mathrm{min}}}^{x_{j-1/2}} u b(u,v) \frac{a(v)}{v}   g_l^0(\tau) \mathrm{d}u \mathrm{d}v \\
& = - \sum_{l=j+1}^N \int_{I_l} \int_{x_{\mathrm{min}}}^{x_{j-1/2}} u b(u,v) \frac{a(v)}{v}   g_l^0(\tau) \mathrm{d}u \mathrm{d}v \\
& \qquad \qquad - \sum_{l=j+1}^N \int_{I_l} \int_{x_{j-1/2}}^{x_{j+1/2}} u b(u,v) \frac{a(v)}{v}   g_l^0(\tau) \mathrm{d}u \mathrm{d}v \\
& \qquad \qquad + g_j^0(\tau) \int_{I_j} \int_{x_{\mathrm{min}}}^{x_{j-1/2}} u b(u,v) \frac{a(v)}{v}  \mathrm{d}u \mathrm{d}v\\
& \qquad \qquad + \sum_{l=j+1}^N \int_{I_l} \int_{x_{\mathrm{min}}}^{x_{j+1/2}} u b(u,v) \frac{a(v)}{v}   g_l^0(\tau) \mathrm{d}u \mathrm{d}v.
\end{aligned}
\label{eq:dF_part1}
\end{equation}
By simplifying Eq.~\ref{eq:dF_part1},
\begin{equation}
\begin{aligned}
& F_{\mathrm{frag}}(x_{j+1/2},\tau)-F_{\mathrm{frag}}(x_{j-1/2},\tau) \\
& = g_j^0(\tau) \int_{I_j} \int_{x_{\mathrm{min}}}^{x_{j-1/2}} u b(u,v) \frac{a(v)}{v}  \mathrm{d}u \mathrm{d}v \\
& \qquad \qquad - \sum_{l=j+1}^N \int_{I_l} \int_{x_{j-1/2}}^{x_{j+1/2}} u b(u,v) \frac{a(v)}{v}   g_l^0(\tau) \mathrm{d}u \mathrm{d}v.
\end{aligned}
\end{equation}
Eq.~\ref{eq:EF} writes
\begin{equation}
\begin{aligned}
g_j^{0,n+1} &=  g_j^{0,n}\left( 1-\frac{\Delta \tau}{\Delta x_j} \int_{I_j} \int_{x_{\mathrm{min}}}^{x_{j-1/2}} u b(u,v) \frac{a(v)}{v}  \mathrm{d}u \mathrm{d}v \right) \\
& \qquad + \frac{\Delta \tau}{\Delta x_j} \sum_{l=j+1}^N \int_{I_l} \int_{x_{j-1/2}}^{x_{j+1/2}} u b(u,v) \frac{a(v)}{v}   g_l^0(\tau) \mathrm{d}u \mathrm{d}v\\
& \geq g_j^{0,n}\left( 1-\frac{\Delta \tau}{\Delta x_j} \int_{I_j} \int_{x_{\mathrm{min}}}^{x_{j-1/2}} u b(u,v) \frac{a(v)}{v}  \mathrm{d}u \mathrm{d}v \right).
\end{aligned}
\end{equation}
Therefore, to guarantee the positivity of $g_j^{0,n+1}>0$ at time-step $n+1$, the CFL condition writes
\begin{equation}
\Delta \tau \; \underset{j}{\mathrm{sup}} \left( \frac{1}{\Delta x_j} \int_{I_j} \int_{x_{\mathrm{min}}}^{x_{j-1/2}} u b(u,v) \frac{a(v)}{v}  \mathrm{d}u \mathrm{d}v \right) <1.
\end{equation}

\end{document}